\documentclass[iop,apj,tighten]{emulateapj}
\usepackage{color}
\usepackage{epsfig}
\usepackage{rotating}
\usepackage{amsmath}
\usepackage{footnote}
\usepackage{setspace}
\usepackage{courier}

\newcommand{\beq}{\begin{equation}}
\newcommand{\eeq}{\end{equation}}
\newcommand{\beqa}{\begin{eqnarray}}
\newcommand{\eeqa}{\end{eqnarray}}

 \def\fun#1#2{\lower3.6pt\vbox{\baselineskip0pt\lineskip.9pt
        \ialign{$\mathsurround=0pt#1\hfill##\hfil$\crcr#2\crcr\sim\crcr}}}

\begin{document}

\title{The Effect of Fiber Collisions on the Galaxy Power Spectrum Multipoles} 

\author{ChangHoon~Hahn\altaffilmark{1}, 
Roman~Scoccimarro\altaffilmark{1}, 
Michael R.~Blanton\altaffilmark{1}, 
Jeremy L.~Tinker\altaffilmark{1}
Sergio Rodr\'{i}guez-Torres\altaffilmark{2,3,4}} 
\altaffiltext{1}{Center for Cosmology and Particle Physics, Department of Physics, New York University, 4 Washington Place, New York, NY 10003; chh327@nyu.edu}
\altaffiltext{2}{Instituto de F\'{i}sica Te\'{o}rica, (UAM/CSIC), Universidad Aut\'{o}noma de Madrid, Cantoblanco, E-28049 Madrid, Spain}
\altaffiltext{3}{Campus of International Excellence UAM+CSIC, Cantoblanco, E-28049 Madrid, Spain}
\altaffiltext{4}{Departamento de F\'{i}sica Te\'{o}rica M8, Universidad Aut\'{o}noma de Madrid, Cantoblanco, E-28049, Madrid, Spain}

\begin{abstract}
\qquad Fiber-fed multi-object spectroscopic surveys, with their ability to collect an unprecedented number of redshifts, currently dominate large-scale structure studies. However, physical constraints limit these surveys from successfully collecting redshifts from galaxies too close to each other on the focal plane. This ultimately leads to significant systematic effects on galaxy clustering measurements. Using simulated mock catalogs, we demonstrate that fiber collisions have a significant impact on the power spectrum, $P(k)$, monopole and quadrupole that exceeds sample variance at scales smaller than $k\sim0.1~h/{\rm Mpc}$.

\qquad We present two methods to account for fiber collisions in the power spectrum. The first, statistically reconstructs the clustering of fiber collided galaxy pairs by modeling the distribution of the line-of-sight displacements between them. It also properly accounts for fiber collisions in the shot-noise correction term of the $P(k)$ estimator. Using this method, we recover the true $P(k)$ monopole of the mock catalogs with residuals of $<0.5\%$ at $k=0.3~h/{\rm Mpc}$ and $<4\%$ at $k=0.83~h/{\rm Mpc}$ -- a significant improvement over existing correction methods. The quadrupole, however, does not improve significantly.

\qquad The second method models the effect of fiber collisions on the power spectrum as a convolution with a configuration space top-hat function that depends on the physical scale of fiber collisions. It directly computes theoretical predictions of the fiber-collided $P(k)$ multipoles and reduces the influence of smaller scales to a set of nuisance parameters. Using this method, we reliably model the effect of fiber collisions on the monopole and quadrupole down to the scale limits of theoretical predictions. The methods we present in this paper will allow us to robustly analyze galaxy power spectrum multipole measurements to much smaller scales than previously possible.

\end{abstract}


\section{Introduction} 
Cosmological measurements such as galaxy clustering statistics are
no longer dominated by uncertainties from statistical precision, but from 
systematic effects of the measurements. This is a result of the millions of 
redshifts to distant galaxies that have been obtained through redshift surveys
such as the 2dF Galaxy Redshift Survey (2dFGRS; \citealt{Colless:1999aa}) and 
the Sloan Digital Sky Survey III Baryon Oscillation Spectroscopic Survey 
(SDSS-III BOSS; \citealt{Anderson:2012aa, Dawson:2013aa}). Current surveys, 
such as the Extended Baryon Oscillation Spectroscopic Survey (eBOSS; \citealt{Dawson:2015aa}), 
and future surveys such as the Dark Energy Survey Instrument (DESI; \citealt{Schlegel:2011aa, 
Morales:2012aa, Makarem:2014aa}), and the Subaru Prime Focus Spectrograph 
(PFS; \citealt{Takada:2014aa}), 
will continue to collect many more million redshifts, extending our measurements 
to unprecedented statistical precision. These completed and future surveys, all use 
and will use fiber-fed spectrographs. 


For each galaxy, a fiber is used to obtain a spectroscopic redshift. However, 
the physical size of the fiber housing and other physical constraints limit 
how well any of these surveys can observe close pairs of galaxies. In the SDSS, 
if two galaxies are located within the fiber collision angular scale from 
one another on the sky, separate fibers cannot be placed adjacently to 
observe them simultaneously 
(\citealt{Yoon:2008aa}). In these situations, only a single redshift 
is measured. With redshifts of galaxies in close angular proximity missing 
from the sample, any clustering statistic probing these scales will be 
systematically affected. 

As our cosmological surveys extend further to higher redshifts, the systematic
effect becomes more severe. The fiber collision angular scale corresponds 
to a larger comoving scale at higher redshift, thereby affecting our measurements on larger 
scales. BOSS, in particular, has an angular fiber collision scale 
of $62\arcsec$. This corresponds to $\sim 0.43 \;\mathrm{Mpc}/h$ at the 
center of the survey's redshift range; fiber-collided galaxies 
account for $\sim 5\%$ of the galaxy sample (\citealt{Anderson:2012aa, 
Reid:2012aa, Guo:2012aa}). 
While this may seem like a relatively small fraction of redshifts, its 
effect on clustering measurements such as the power spectrum and bispectrum 
is significant and needs to be accounted for in order to probe mildly non-linear scales. 
Unfortunately, future spectroscopic surveys such 
as DESI, which will use robotic fiber positioner 
technology, will be subject to similar effects. Therefore, 
accounting for the effects of fiber collisions will remain a crucial and 
unavoidable challenge for analyzing clustering measurements. 

%

To correct for fiber collisions, one common approach used in clustering 
measurements is the nearest neighbor method (\citealt{Zehavi:2002aa, Zehavi:2005aa, 
Zehavi:2011aa, Berlind:2006aa, Anderson:2012aa}). For fiber-collided galaxies without 
resolved redshifts, the method assigns the statistical weight of the 
fiber-collided galaxy to its nearest angular neighbor. This provides 
a reasonable correction for the fiber collision effects at scales much 
larger than the fiber collision scales; however the correction falls short 
elsewhere. In fact, as \cite{Zehavi:2005aa} find, 
fiber collisions affect the two-point correlation function (2PCF) 
measurements even on scales significantly larger than the fiber collision 
scale ( $> 1\;\mathrm{Mpc}/h$). 

For power spectrum measurements in BOSS, 
the nearest neighbor method has recently been supplemented with adjustments 
in the constant shot-noise term of the power spectrum estimator to correct 
for fiber collisions~\citep{Beutler:2014aa, Gil-Marin:2014aa, Gil-Marin:2015aa, 
Gil-Marin:2016ab, Beutler:2016aa, Grieb:2016aa, Gil-Marin:2016aa}. More specifically, 
methods like the one used in 
\cite{Gil-Marin:2014aa} obtain the value of the shot-noise term from mock catalogs and thus rely entirely on their accuracy 
to correct for fiber collisions. This is concerning since, as we shall demonstrate in detail, fiber 
collisions depend systematically on the small-scale power spectrum, and mock catalogs used for large scale structure analyses are typically 
not based on high resolution N-body simulations. In addition, there is no way to validate and 
calibrate the shot-noise term independently for observations. A more 
reliable approach is to marginalize over the value of the shot-noise 
term, and this is the approach that has recently become more popular
~\citep{Beutler:2014aa, Gil-Marin:2016ab, Beutler:2016aa, Grieb:2016aa, 
Gil-Marin:2016aa}. However, adjustments 
to the shot-noise term are limited to the power spectrum monopole, since higher 
order multipoles do not have a shot-noise term. However, as we shall discuss in detail below, {\em fiber collisions
affect all multipoles in a $k$-dependent way}, not just adding a constant for the monopole power.

\cite{Guo:2012aa}, focusing on SDSS-III BOSS like samples, proposed 
a fiber collision correction method for the 2PCF that is able to reasonably 
correct for fiber collisions above and below the collision scale. 
\cite{Guo:2012aa} estimates the total contribution of fiber-collided galaxies 
to the 2PCF by examining the pair statistics in overlapping tiling regions of 
the survey, where a smaller fraction of galaxies suffer from fiber collisions.
Unfortunately, applying an analogous method in Fourier space proves to be more difficult. 
The \cite{Guo:2012aa} method in Fourier space would involve measuring the power spectra 
for individual overlapping regions. Given the complex geometry of these 
regions, the systematic effect introduced by the window function makes 
measuring the power spectrum at larger scales intractable. 


Meanwhile, galaxy redshift-space power spectrum models from perturbation theory continue to 
reliably model higher $k$ in the weakly non-linear regime
\citep{Taruya:2010aa, Sato:2011aa, Taruya:2012aa, Okumura:2012aa, Taruya:2013aa,
Taruya:2014aa, Beutler:2014aa, Okumura:2015aa, Beutler:2016aa, Grieb:2016aa, Sanchez:2016aa}.
Recent analyses of galaxy power spectrum 
multipoles (\citealt{Zhao:2013aa, Beutler:2014aa, Gil-Marin:2014aa,
Gil-Marin:2016ab, Beutler:2016aa, Grieb:2016aa, Gil-Marin:2016aa}) use scales up to $k_{\rm max}=0.15-0.2 h$/Mpc for BOSS galaxies, and this limit will for sure move towards smaller scales in upcoming analyses. As statistical errors decrease the importance of systematics due to fiber collisions plays an increasingly important role. The main goal of this paper is to quantify this systematic effect for the power spectrum multipoles and to provide ways to overcome it; for this purpose we
develop two distinct approaches. 

The first approach improves upon the nearest neighbor method by modeling the 
distribution of the line-of-sight displacement between resolved fiber collided 
galaxies to statistically reconstruct the clustering of fiber-collided galaxies. 
This uses information on resolved fiber collided galaxies that is available from 
the data themselves (e.g. in tiling overlap regions). The difficulty with this 
method is that it works statistically, i.e. we cannot reconstruct the {\em actual} 
galaxy by galaxy line of sight displacement due to collisions. As a result of this, 
while the method works very well to recover the true power spectrum monopole 
from fiber collided galaxy catalogs, it does not work sufficiently well for the 
power spectrum quadrupole which is far more sensitive to the precise structure of ``fingers of god''. 

The second approach addresses the shortcomings of the first one by modeling the effects of fiber collisions on the {\em predictions} instead of trying to undo their effect on the data before computing power spectrum statistics. It approximates the effect of fiber collisions on the 2PCF as 
a 2D top hat function. Then it derives the effect of fiber collisions on the galaxy power spectrum as a 
convolution of the true power spectrum with the top hat function. Therefore the theoretical predictions for the power spectrum are fiber collided and then can be compared  directly to the observed fiber collided power spectrum in clustering analyses. 

This paper is organized as follows. 
In Section \ref{sec:catalog}, we briefly describe the simulated mock catalogs 
with realistic fiber collisions and the power spectrum estimator used throughout 
the paper. We then demonstrate the impact of fiber collisions on power spectrum 
measurements and how the nearest neighbor method does not adequately account 
for fiber collisions in Section \ref{sec:fc_pk}. We present our two methods 
of accounting for fiber collisions along with the results for mock catalogs in 
Section \ref{sec:dlospeak} and Section \ref{sec:fourier}, respectively. Finally 
in Section \ref{sec:summary} we summarize our results and conclude. 

\section{Fiber-collided Mock catalogs} \label{sec:catalog}
For various purposes, such as characterizing the impact of the survey window function on statistics and estimating covariance matrices, simulated mock 
catalogs play a crucial role in interpreting  
clustering measurements of observed galaxies 
\citep{Cole:1998aa, Scoccimarro:2002aa, Yan:2004aa, Anderson:2012aa, Manera:2013aa, 
Monaco:2013aa, Beutler:2014aa, Gil-Marin:2014aa, White:2014aa, Manera:2015aa,
Tassev:2015aa, Carretero:2015aa, Howlett:2015aa, Izard:2016aa, Chuang:2015aa,
Kitaura:2016aa, Munari:2016aa, Sunayama:2016aa}. 
They also provide a means of understanding systematic effects such as  
fiber collisions (\citealt{Guo:2012aa, Manera:2013aa}).
Since systematic effects can be simulated on them, they allow us to test 
how these effects influence clustering measurements and devise correction 
methods that attempt to account for these effects.

A direct way of understanding the effects of fiber collisions on clustering 
statistics in observations is to first apply fiber collisions to mock catalogs
and then compare the clustering statistics obtained from mock catalogs with 
and without the fiber collisions. Correction methods for fiber collisions can 
then be applied to the fiber-collided mocks. The merit of the correction 
method can be assessed by how successfully they reproduce the clustering statistics 
of the original mock catalogs without fiber collisions. The correction 
method can then be applied to the observed data with some assurance that it 
accounts for fiber collisions and improves the clustering measurements. 

When applying the fiber collisions to the mock catalogs, it is essential to 
apply them in the same manner they affect the observations. For BOSS, galaxies
within $62\arcsec$ are fiber-collided (\citealt{Anderson:2012aa}). In reality, 
this criteria is further complicated by the tiling scheme of observing 
plates that create overlapping regions, which have a higher success rate in 
resolving galaxy spectra within the fiber collision angular scale (\citealt{
Guo:2012aa, Reid:2012aa}). Furthermore, fiber collisions are only one of the 
systematic effects that influence BOSS data. Systematic effects include the 
unique geometry of the BOSS survey, the variable completeness in different areas  
covered by unique sets of spectroscopic plates, and redshift failures 
(\citealt{Anderson:2012aa, Ross:2012aa}). 

\def \cmasscolor{black}
\def \ldgcolor{blue}
\def \nseriescolor{orange}
\def \qpmcolor{blue}
\def \tmcolor{green}
\def \bmdcolor{red}

\begin{figure}
\begin{center}
\includegraphics[scale=0.425]{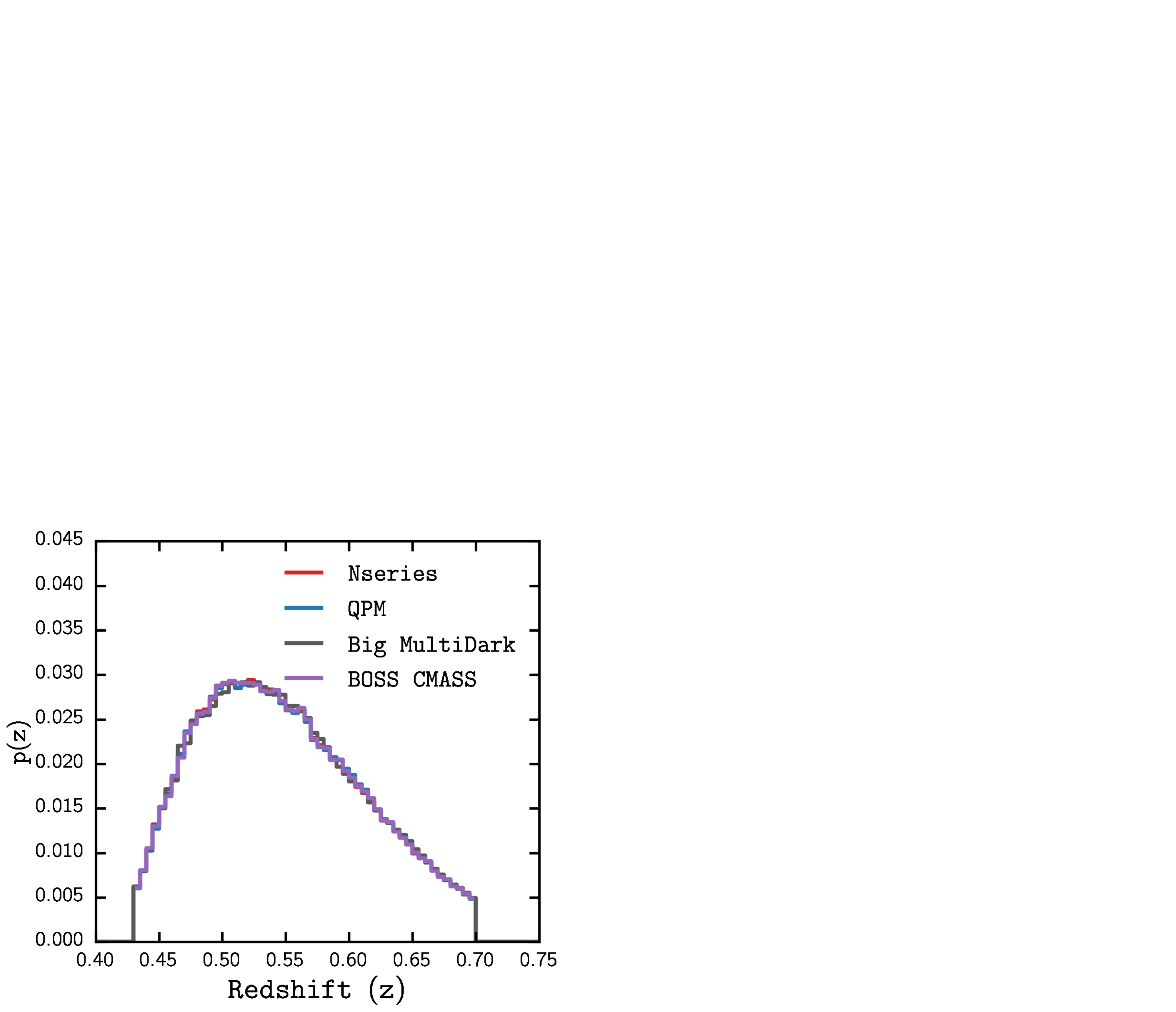} 
\caption{Normalized galaxy redshift distribution of the Nseries (\nseriescolor), 
QPM (\qpmcolor), and BigMultiDark (\bmdcolor) mock catalogs. The 
normalized redshift distribution of BOSS DR12 CMASS sample galaxies 
is also plotted (\cmasscolor). Each of the distributions were computed
with a bin size of $\Delta z = 0.025$. All of the mock catalogs used 
in this work closely trace the BOSS CMASS redshift distribution.}
\label{fig:zdist}
\end{center}
\end{figure}

Effects of fiber collisions must be understood and interpreted in conjunction 
with the other systematic effects. Therefore, in this paper, we use Quick 
Particle Mesh (\citealt{White:2014aa}), Nseries (Tinker et al. in prep), and 
the BigMultiDark (\citealt{Rodriguez-Torres:2015aa}) mock catalogs, which 
have already been extensively used in interpreting clustering results for  
BOSS and are generated through different prescriptions. Therefore they 
provide a robust sets of data to measure the effects of fiber collisions and 
to test our correction methods. 

The QPM mock galaxy catalogs uses a ``quick particle mesh" method, which uses 
a low resolution particle-mesh N-body solver, with a resolution of 
$2\;\mathrm{Mpc}/h$, to evolve particles within a 
periodic simulation volume. The particles are assigned halo masses in order 
to match the halo mass function and large-scale bias of halos of high resolution 
simulations. Afterwards the HOD parameterization of \cite{Tinker:2012aa} is 
used to populate the halos. The mock galaxy sample is then trimmed to the 
BOSS CMASS survey footprint, downsampled based on angular sky completeness 
(sector completeness) and radial selection. Furthermore, QPM mocks model the 
fiber collisions of the BOSS CMASS sample ($62\arcsec$). QPM uses the following 
$\Lambda$CDM cosmology: $\Omega_\mathrm{m} = 0.29$, $\Omega_\Lambda = 0.71$, 
$\sigma_8 = 0.8$, $n_\mathrm{s} = 0.97$ and $h=0.7$. We use 100 realizations 
of the QPM catalog. For a detailed description of the QPM galaxy mock catalogs 
we refer readers to \cite{White:2014aa}. 

Next, the Nseries mock catalogs are created from a series of high-resolution 
N-body simulations. Each mock has the same angular selection function as the 
North Galactic Cap region of the BOSS DR12 large-scale structure sample for 
CMASS galaxies (\citealt{Cuesta:2016aa}). They also reproduce the 
redshift distribution of the BOSS CMASS sample. The Nseries mock catalogs are 
created from seven independent N-body simulations, each of the same cosmology. 
Each simulation box is $2.5\;\mathrm{Gpc}/h$ per side with cosmology: 
$\Omega_\mathrm{m} = 0.286$, $\Omega_\Lambda = 0.714$, $\sigma_8 = 0.82$,
$n_\mathrm{s} = 0.96$ and $h=0.7$. Out of these Nseries box simulations, the 
three orthogonal projections of each box is used to create $84$ mocks.
Each of the cut-out mocks is then passed through the same fiber assignment 
code as the actual BOSS data using the distribution of plates in BOSS. 
Thus, the angular variation of fiber collisions faithfully reproduces 
that of the data, with $\sim 5\%$ of the targets without fibers due to close 
neighbors in regions of the footprint only covered by one tile. 

\begin{figure*}
\begin{center}
\includegraphics[scale=0.55]{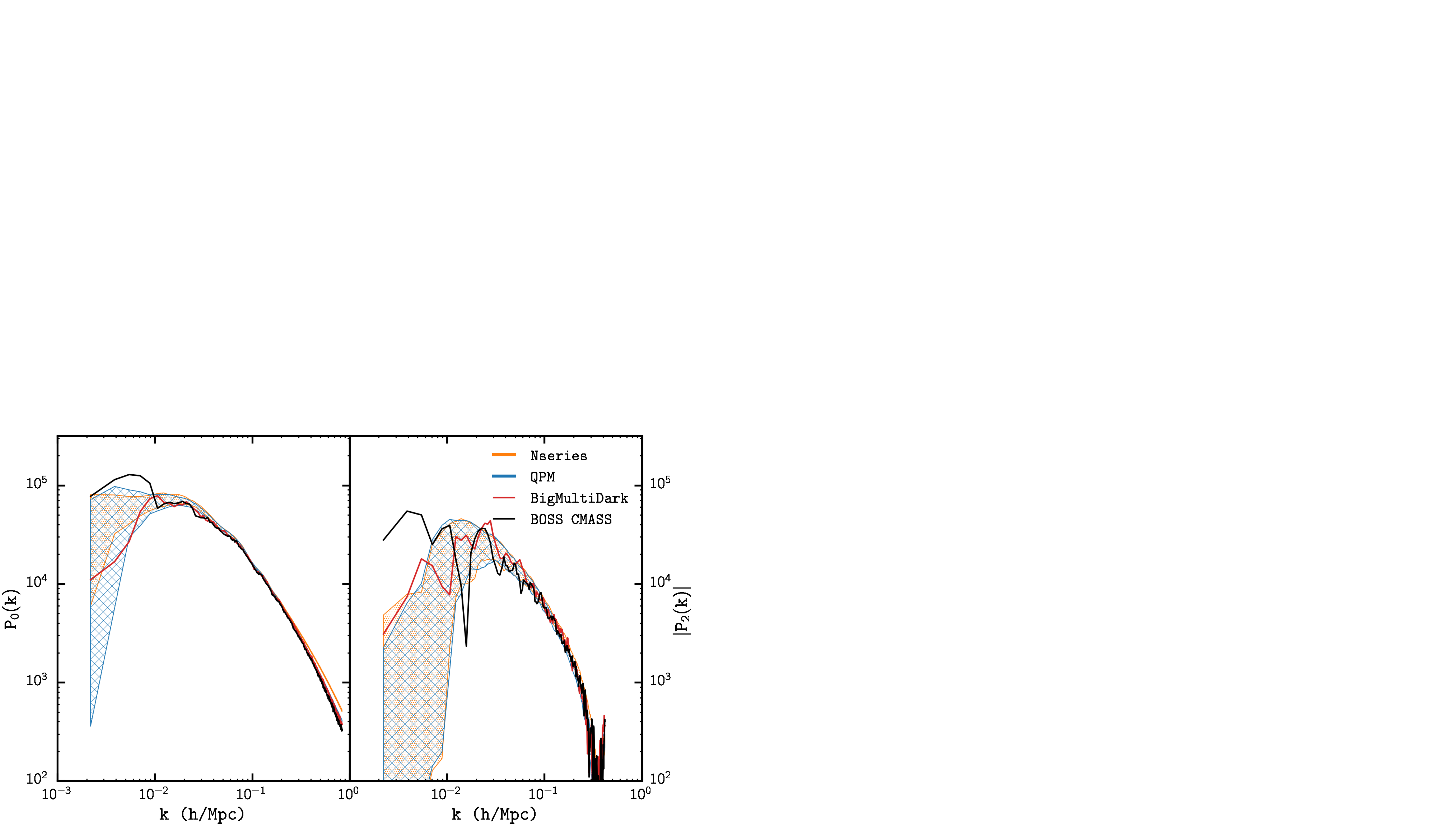} 
\caption{
Power spectrum monopole $P_0(k)$ and 
quadrupole $|P_2(k)|$ measurements for the Nseries (\nseriescolor), 
QPM (\qpmcolor), and BigMultiDark (\bmdcolor) mock catalogs 
(Section \ref{sec:catalog}). The $P_l(k)$ measurements 
for the Nseries and QPM mock catalogs are averaged over the
multiple mock realizations and the width of the power spectra represents 
the sample variance ($\sigma_l(k)$; Eq.~\ref{eq:pk_var}) of the realizations. 
For the quadrupole, we plot the $|{P_2(k)}|$ instead of 
${P_2(k)}$ because the measurement becomes negative for 
$k \gtrsim 0.35\;h/\mathrm{Mpc}$. For comparison, we also include the monopole and 
quadrupole power spectra of the BOSS DR12 CMASS sample, which are calculated 
using the same estimator but with statistical weights described in Eq.~(\ref{eq:weight}). 
While fiber collisions are inevitably included in the BOSS CMASS power spectra, 
they are {\it not} yet applied to the mock catalogs power spectra measurements above. } 
\label{fig:mockpk}
\end{center}
\end{figure*}

Finally the BigMultiDark galaxy mock catalog is generated using the 
BigMultiDark Planck (BigMDPL), one of the MultiDark3 N-body simulations 
(\citealt{Klypin:2014aa}). BigMDPL uses a GADGET-2 code (\citealt{Springel:2005aa})
in a cubic box of $2.5\;h^{-1}\mathrm{Gpc}$ sides with $3840^3$ dark matter 
particles and a mass resolution of $2.4\times 10^{10} h^{-1}M_\odot$. 
As the name suggests, BigMDPL uses Planck cosmological parameters in a flat $\Lambda$CDM cosmology: 
$\Omega_m = 0.307$, $\Omega_B = 0.048$, $\Omega_\lambda = 0.693$, $\sigma_8 = 0.829$, 
$n_s = 0.96$ and $h = 0.678$. 

From the BigMDPL N-body simulation, \cite{Rodriguez-Torres:2015aa} uses 
the $\mathtt{RockStar}$ (Robust Overdensity Calculation using K-Space Topologically 
Adaptive Refinement) halo finder (\citealt{Behroozi:2013aa}) to obtain 
a dark matter halo catalog. Afterwards, they use the 
SUrvey GenerAtoR code ($\mathtt{SUGAR}$) to generate a galaxy catalog from the halo 
catalog. $\mathtt{SUGAR}$ uses halo abundance matching with an intrinsic scatter 
on the stellar mass function of the Portsmouth SED-fit DR12 stellar mass
catalog (\citealt{Maraston:2013aa}) to populate the dark matter halos with  
galaxies. \cite{Rodriguez-Torres:2015aa} then model fiber collisions using 
\cite{Guo:2012aa} in order to reproduce the effect of fiber collisions on 
the observed BOSS galaxies. For any further details on the BigMultiDark 
galaxy mock catalog, we refer readers to \cite{Rodriguez-Torres:2015aa}.

In Figure \ref{fig:zdist}, we plot the normalized redshift distribution of the 
Nseries (\nseriescolor), QPM (\qpmcolor), and BigMultiDark (\bmdcolor) 
mock catalogs along with the redshift distribution of the BOSS DR12 
CMASS sample galaxies. All of these mock catalogs were constructed
for the BOSS analysis and their redshift distributions closely 
trace the observed BOSS distribution.

\subsection{Power Spectrum Estimator} \label{sec:pk_est}
In this paper, out of the many possible clustering measurements, we focus on the 
galaxy power spectrum and its monopole and quadrupole in redshift space. 
Throughout the paper, unless specified, when we measure the 
power spectrum we use the estimator described in \cite{Scoccimarro:2015aa}, 
which accounts for radial redshift space distortions (see also \citealt{Bianchi:2016aa}). 
In this estimator, galaxies are interpolated and Fast Fourier 
transformed as discussed in \cite{Sefusatti:2016aa}. 
Since the algorithm is efficient, it makes power spectrum computations 
for large number of mock realizations tractable.

To summarize the method, we calculate the 
monopole component of the power spectrum using: 
\begin{equation}\label{eq:roman_p0k}
\widehat{P_0}(k) = \frac{1}{I_{22}} \left[ \int \frac{d\Omega_k}{4 \pi} |F_0({\bf k})|^2 - N_0 \right]
\end{equation}
where 
\begin{equation}
F_0({\bf k}) = \left( \sum_{j = 1}^{N_g} - \alpha \sum_{j=1}^{N_r} \right) w_j\;e^{i {\bf k}\cdot{\bf x}_j}
\end{equation}
with normalization constant
\begin{equation} \label{eq:i22}
I_{22} = \alpha \sum^{N_r}_{j=1} \bar{n}({\bf x}_j)w_j^2
\end{equation}
and shot noise term following from the estimator is \citep{Scoccimarro:2015aa}
\begin{equation} \label{eq:roman_shotnoise}
N_0 = \left( \sum_{j = 1}^{N_g} + \alpha^2 \sum_{j=1}^{N_r} \right) w_j^2, 
\end{equation}
which represents the constant shot noise contribution to the power due to 
the discrete density field of our galaxies and random catalog. Here $\alpha$ is the ratio of the 
number of galaxies ($N_g$) over the number of synthetic random galaxies 
($N_r$), $\bar{n}({\bf x})$ is the mean density of the galaxies at position 
${\bf x}$, and $w_j$ is weight of each object, which includes the minimum 
variance weight from \cite{Feldman:1994aa}: 
\begin{equation}
w_{\mathrm{FKP}} ({\bf x}_j) = \frac{1}{1+\bar{n}({\bf x}_j) P_0}
\end{equation}
where $P_0$ is the power spectrum amplitude at which the error is minimized. 
We use $P_0 = 20000\; \mathrm{Mpc}^3/h^3$ for our analysis, which corresponds 
to $k \sim 0.1\; h/\mathrm{Mpc}$. We note that the shot noise term in 
Eq.~(\ref{eq:roman_shotnoise}) differs from the standard shot noise term from
\cite{Feldman:1994aa}. The difference between various shot noise expressions used in the literature  
will be discussed in detail in Section \ref{sec:shotnoise}. 

For the quadrupole, we have 
\begin{equation}\label{eq:roman_p2k}
\widehat{P_2}(k) = \frac{5}{I_{22}} \int \frac{d\Omega_k}{4 \pi} F_2({\bf k}) F_0^*({\bf k})
\end{equation}
where 
\begin{equation}
F_2({\bf k}) = \frac{3}{2}\hat{k}_a\hat{k}_b Q^{ab}({\bf k}) - \frac{1}{2} F_0({\bf k})
\end{equation}
with 
\begin{equation}
Q^{ab}({\bf k}) = \left( \sum_{j = 1}^{N_g} - \alpha \sum_{j=1}^{N_r} \right) \hat{x}_j^a\hat{x}_j^b w_j \;e^{i {\bf k}\cdot{\bf x}_j}
\end{equation}

In Figure \ref{fig:mockpk}, we plot the  power spectrum monopole and quadrupole, ${P_0(k)}$ 
and $|{P_2(k)}|$, measured using Eq.~(\ref{eq:roman_p0k}) and Eq.~(\ref{eq:roman_p2k}), respectively, for the Nseries, QPM, and BigMultiDark 
mock catalogs. We plot $|{P_2(K)}|$ because the power spectrum quadruple 
becomes negative for $k \gtrsim 0.35\;h/\mathrm{Mpc}$. ${P_0(k)}$ and
$|{P_2(k)}|$ are averaged over the $84$ and $100$ realizations for
Nseries and QPM. We note that fiber collisions are not applied to these mock 
catalogs. Without fiber collisions, the weights of the objects are equivalent 
to the FKP weights, $w_j = w_{j,\mathrm{FKP}}$. 

We also plot the $P_0(k)$ and $P_2(k)$ of the BOSS Data Release 12 CMASS 
data (\cmasscolor) in Figure \ref{fig:mockpk}. For BOSS DR12 CMASS, 
systematic weights are assigned to the galaxies in order to account for 
sector completeness, redshift failures, and fiber collisions. Each galaxy 
has a statistical weight determined by, 
\begin{equation} \label{eq:weight}
w_{j, \mathrm{tot}} = w_{j, \mathrm{sys}} (w_{j, \mathrm{rf}} + w_{j, \mathrm{fc}} -1), 
\end{equation} 
(\citealt{Anderson:2012aa, Ross:2012aa, Beutler:2014aa}), which are included in the 
final object weight $w_j$ along with $w_{j, \mathrm{FKP}}$. In this formula, $w_{j,\mathrm{rf}}$ is a weight that accounts for redshift failures and $w_{j,\mathrm{fc}}$
is the fiber collision weight determined by the nearest angular neighbor method, which we
later discuss in Section \ref{sec:fc_pk}. The statistical 
weights are also included in $\alpha = \sum_{j=1}^{N_g} w_\mathrm{tot} / N_r$. 
We note that fiber collisions are inevitably included in the CMASS $P_l(k)$.
However they are not yet included in the $P_l(k)$ of the mock catalogs in Figure \ref{fig:mockpk}.

For the mock catalogs with multiple realizations (QPM and Nseries), we 
compute the sample variance of the power spectrum
\begin{equation} \label{eq:pk_var}
\sigma_l (k)= \sqrt{\frac{1}{N_\mathrm{mocks}-1} \sum\limits_{i=1}^{N_\mathrm{mocks}} (P^i_l(k)- \langle{P_l(k)}\rangle)^2 \ }. 
\end{equation}
$N_\mathrm{mock}$ is the number of mock realizations (84 for Nseries and 100 for 
QPM) and $P^i_l(k)$ is the power spectrum for each realization. $\sigma_l(k)$ is 
represented in Figure \ref{fig:mockpk} by the width of the shaded regions. 
\begin{figure*}
\begin{center}
\includegraphics[scale=0.55]{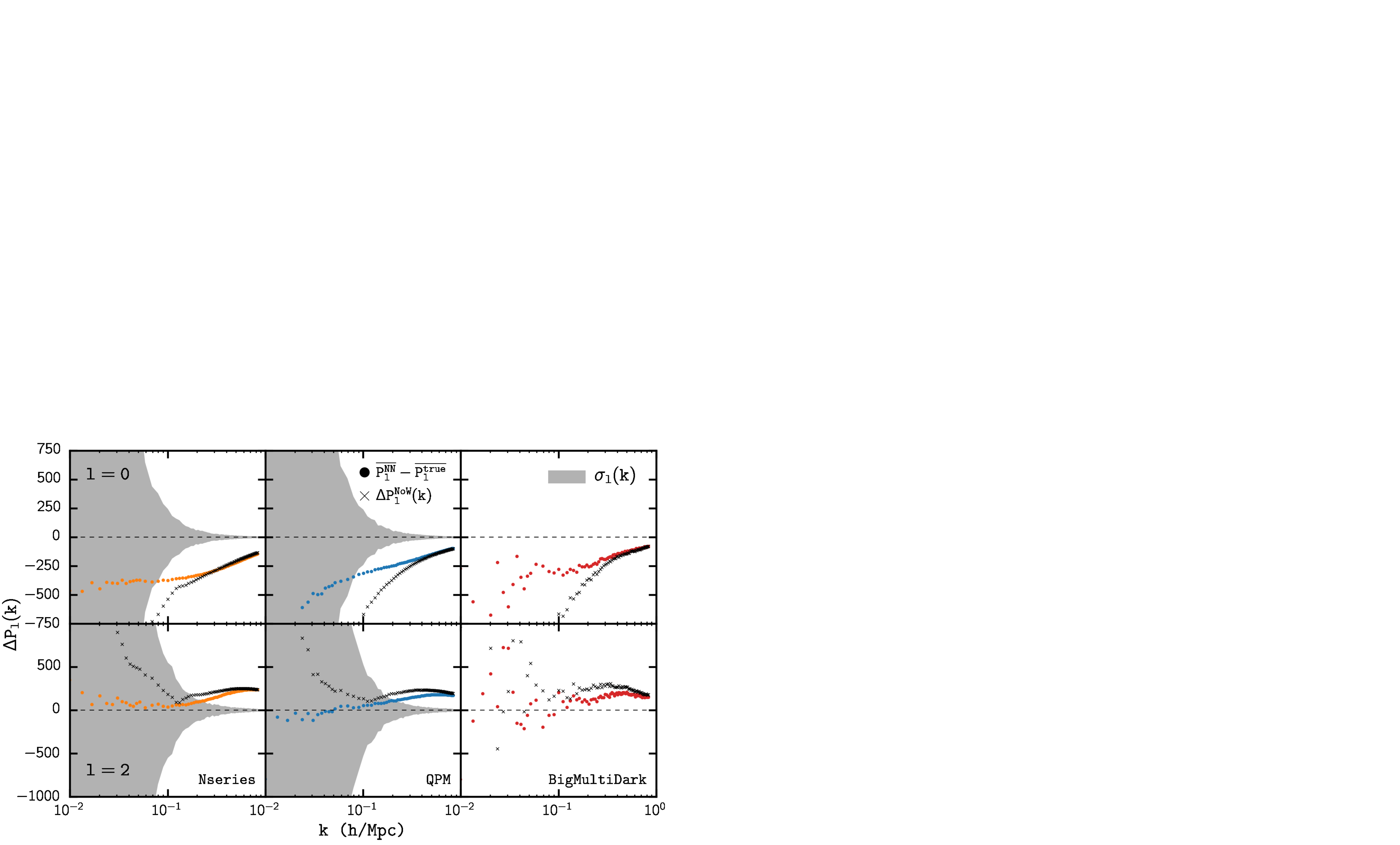} 
\caption{The fiber collision power spectrum residual, 
$(P_l^\mathrm{NN}-P_l^\mathrm{true})$ 
(Section \ref{sec:fc_pk}), for the monopole (top) and quadrupole 
(bottom) of the Nseries (left), QPM (middle), and 
BigMultiDark (right) mock catalogs. For the Nseries and QPM mocks, 
we plot the sample variances $\sigma_l(k)$ (grey shaded region) of 
$P_l^\mathrm{true}(k)$ for comparison. The power spectrum residual for the 
NN method is an improvement over the residual with no correction 
($\Delta P_l^\mathrm{NoW}(k)$; x) at most scales probed. 
However, we highlight that at $k > 0.1 \;h/\mathrm{Mpc}$ and 
$k > 0.2\;h/\mathrm{Mpc}$, for the monopole and quadrupole 
respectively, the residuals from fiber collision surpass the sample 
variance. 
At smaller scales, NN method does not sufficiently account for 
the effects of fiber collisions in $P_l(k)$ measurements.}
\label{fig:fc_pk}
\end{center}
\end{figure*}

\begin{figure}
\begin{center}
\includegraphics[scale=0.4]{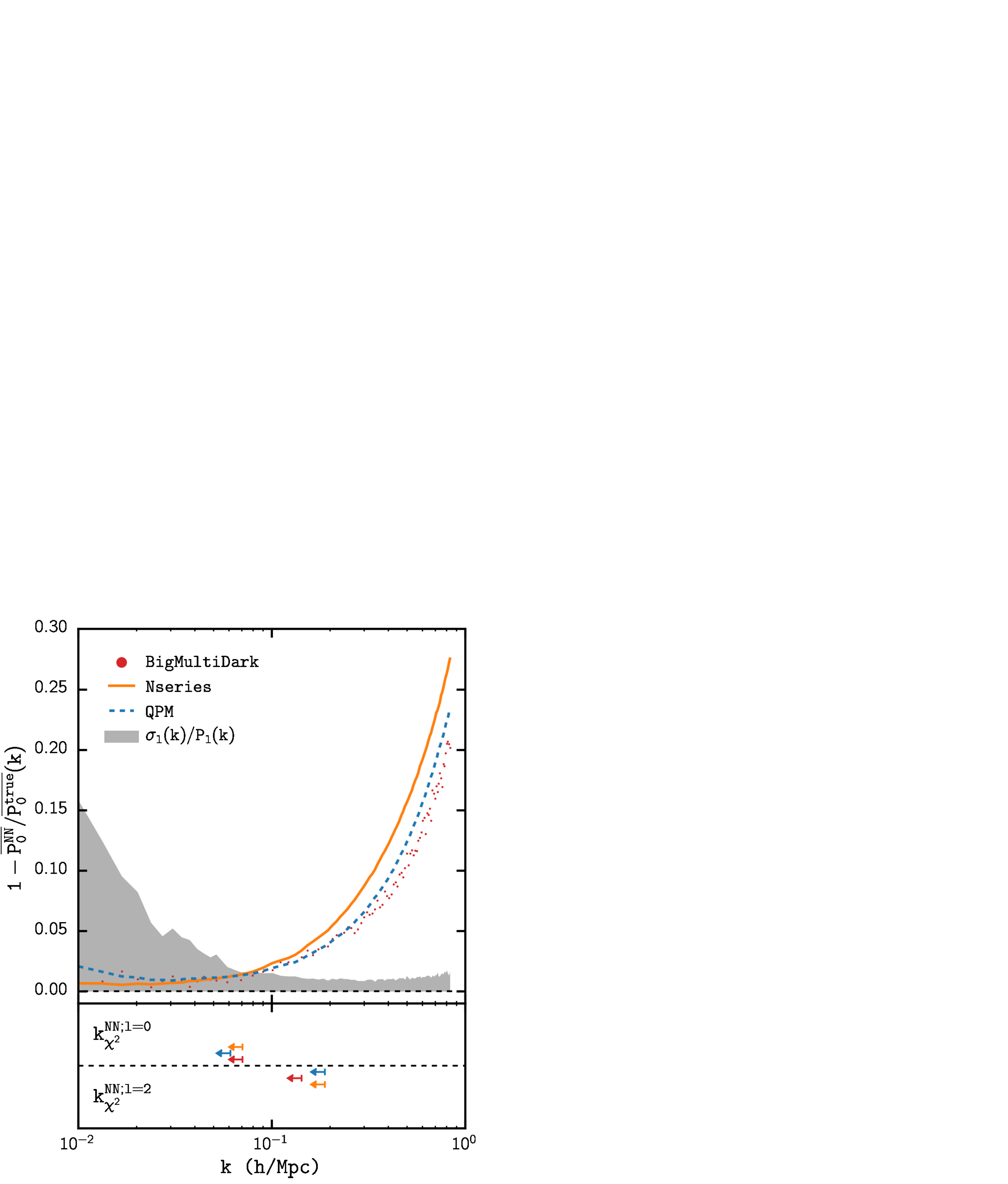} 
\caption{{\it Top Panel}: The normalized residuals, 
$1 - \overline{P_0^\mathrm{NN}}/\overline{P_0^\mathrm{true}}(k)$, 
of the NN method for the Nseries (\nseriescolor), 
QPM (\qpmcolor), and BigMultiDark (\bmdcolor) power spectrum monopole. 
We also plot the normalized sample variance $\sigma_0(k) / P_0(k)$ 
(gray shaded region) of the Nseries mocks for comparison.
The QPM $\sigma_0(k) / P_0(k)$ is effectively the same as the Nseries 
$\sigma_0(k)/P_0(k)$, so we do not included in the figure. 
The comparison reveals that the effect of fiber collisions not only 
biases the power spectrum beyond sample variance at $k \gtrsim 0.1 \;h/\mathrm{Mpc}$, 
but that the effect increases relative to sample variance at smaller scales. At 
$k = 0.2\;h/\mathrm{Mpc}$,  
the normalized residual is greater than $4$ times the normalized sample variance.\\
{\it Bottom Panel}: We mark $k_{\chi^2}$ where $\Delta \chi^2(k_{\chi^2}) = 1$ (Eq.~\ref{eq:chisquared}) for the NN method. $k^\mathrm{NN}_{\chi^2}$ is a conservative 
scale limit of the NN method. Arrows above the dashed 
line mark $k_{\chi^2}$ for the monopole while the arrows below the dashed line mark
$k_{\chi^2}$ for the quadrupole. The color of the arrows indicate the mock catalog: 
Nseries (\nseriescolor), QPM (\qpmcolor), and BigMultiDark (\bmdcolor). Averaged
over the three mock catalogs, we get $k^\mathrm{NN}_{\chi^2} = 0.068$ and $0.17 \;h/\mathrm{Mpc}$.
for the monopole and quadrupole respectively.}
\label{fig:NN_norm_resid}
\end{center}
\end{figure}
\section{Fiber Collision Methods} \label{sec:fc_corr}
\subsection{Nearest Angular Neighbor Method (NN)} \label{sec:fc_pk}
A common approach to accounting for fiber collisions in clustering measurements has been 
to use the nearest angular neighbor method (\citealt{Zehavi:2002aa, Zehavi:2005aa, 
Zehavi:2011aa, Berlind:2006aa, Anderson:2012aa}),
hereafter NN method. For galaxies without resolved spectroscopic redshifts due to fiber 
collisions, the entire statistical weight of the galaxy is assigned to its nearest angular 
neighbor with resolved redshift. This method effectively assumes that all galaxies within 
the angular fiber collision scale ($< 62 \arcsec$ for BOSS) are correlated with one another. 
In the context of the halo model, the NN method assumes that 
galaxies within the fiber collision angular scale reside in the same halo 
so displacing one of the galaxies and placing it on top of the other does 
not significantly impact clustering statistics. This is a reasonable assumption  
for the 2PCF and the power spectrum on scales far greater than fiber collisions.  

One consequence of this method is that galaxies coincidentally within the angular 
fiber collision scale (hereafter referred to as ``chance 
alignments") are incorrectly assumed to be gravitationally correlated and 
within the same halo. So when the statistical weight of the collided galaxy
is added to its nearest angular neighbor, the collided galaxy is in fact 
displaced significantly from its true radial position. This displacement can even 
be on the scale of the survey depth, which corresponds to $\sim500\;\mathrm{Mpc}$
for BOSS. Furthermore, even for fiber collided galaxies that reside in the 
same gravitationally bound structures such as groups or clusters, up-weighting 
the nearest neighbor disregards the line-of-sight displacements within these 
structures. 

To precisely quantify the effect of fiber collisions on the power spectrum, 
we compare the power spectrum measurements of the NN weighted fiber collided
mock catalogs $P_l^\mathrm{NN}$ to the power spectrum measurements of 
the mock catalogs without fiber collisions, the ``true'' power spectrum 
$P_l^\mathrm{true}$. Specifically, in Figure \ref{fig:fc_pk}, we 
plot the power spectrum residual $(P_l^\mathrm{NN} - P_l^\mathrm{true})$ as a function of $k$. 
The power spectrum estimators Eq.~(\ref{eq:roman_p0k}) and~(\ref{eq:roman_p2k})  
are used to calculate the monopole and quadrupole respectively. We include measurements of the sample 
variance, $\sigma_l(k)$, for the Nseries and QPM mock catalogs (Eq.~\ref{eq:pk_var}). 
We also include the power spectrum residual $\Delta P_l^\mathrm{NoW}(k) = 
P_l^\mathrm{NoW}(k) - P_l^\mathrm{true}(k)$ (dashed), where $P_l^\mathrm{NoW}(k)$ is the 
power spectrum of the fiber collided mock catalogs with {\em no} NN weights,
with the collided galaxies removed from the sample. 

As both $P_l(k)$ and $\sigma_l(k)$ vary significantly over the probed $k$ range, the significance 
of the discrepancies between $P^\mathrm{NN}_l(k)$ and $P^\mathrm{true}_l(k)$ 
are not adequately portrayed in Figure \ref{fig:fc_pk}, especially for the monopole. 
Therefore, to compare $P_0^\mathrm{NN}$ and $P_0^\mathrm{true}$ over a wide $k$ 
range and to especially highlight the discrepancies at small scales, in 
Figure \ref{fig:NN_norm_resid}, we compare the normalized monopole residuals, 
$1 - P_0^\mathrm{NN}/P_0^\mathrm{true}$, to the normalized sample variance,
$\sigma_0(k)/P_0^\mathrm{true}$. 

For the monopole, Figure \ref{fig:fc_pk} demonstrates that while the 
NN method (circles) provides an overall improvement over applying no correction (crosses)
at most scales, fiber collisions still significantly bias the corrected 
power spectrum at all scales. The effect also has a significant $k$ 
dependence, which implies that an adjusted 
constant shot noise term alone is insufficient in accounting for the deviation. 
Even at $k \approx 0.1\;h/\mathrm{Mpc}$, the effect of 
fiber collisions in the NN method alarmingly surpasses sample variance. 
While the amplitude of the residual decreases as $k$ increases, Figure
\ref{fig:NN_norm_resid} reveals that as a fraction of 
$P_0^\mathrm{true}(k)$, the discrepancy is in fact increasing. 
In other words, the NN method becomes less effective at correcting for 
fiber collisions on smaller scales, as expected. At the smallest scales probed 
($k = 0.83\;h/\mathrm{Mpc}$), the $P_0^\mathrm{NN}(k)$ underestimates the 
true power spectrum monopole by over $20\%$. 

For the quadrupole, the NN method improves the power spectrum residuals over 
no correction. However, even with the NN method, the effect of fiber collisions begins 
to significantly grow at $k=0.1\;h/\mathrm{Mpc}$ and becomes comparable to 
the sample variance at $k \sim 0.2\;h/\mathrm{Mpc}$. For $k > 0.2\;h/\mathrm{Mpc}$, 
the effect continues to increase and quickly overtakes the decreasing 
sample variance. At the smallest scales measured ($k = 0.83 \;h/\mathrm{Mpc}$) the 
residual is over eight times the sample variance.

Recently power spectrum analyses have measured the power spectrum using a wide range of 
$k$ bins: for example, \cite{Anderson:2012aa} use $\Delta k = 0.04 \;h/\mathrm{Mpc}$ and \cite{Beutler:2014aa} and \cite{Grieb:2016aa} use 
$\Delta k = 0.005\;h/\mathrm{Mpc}$. Here, we use $\Delta k = 0.01\;h/\mathrm{Mpc}$, 
which is within this general range, in agreement with~\cite{Beutler:2016aa} and \cite{Gil-Marin:2016ab}. Sample variance measured with larger $\Delta k$ 
is smaller; so a straight comparison in Figure \ref{fig:fc_pk} between the power spectrum residuals (symbols) 
and the sample variance (shaded region) has a significant dependence on 
the choice of $\Delta k$. What is independent of binning is a cumulative $\chi^2$ as a function of $k$, and thus we define a $k$ scale limit $k_{\chi^2}$  
so that $\Delta  \chi^2(k_{\chi^2}) = 1$, where 
\beq \label{eq:chisquared}
\Delta \chi^2(k') = \sum\limits_{i,j < N_k} 
\left[P^\mathrm{NN}_{l,i} - P^\mathrm{true}_{l,i}\right] C^{-1}_{l;\; i,j}
\left[P^\mathrm{NN}_{l,j} - P^\mathrm{true}_{l,j}\right]
\eeq
where $N_k$ is the number of bins where $k < k'$ and $C^{-1}_{l; i,j}$ are the elements of the
inverse covariance matrix for $P_l^\mathrm{true}(k)$. The elements of the covariance matrix 
$\mathbf{C}_l$ are computed as 
\beq
\mathrm{C}_{l;\; i,j} = \frac{1}{N_{\mathrm{mocks}}-1}\sum_{k=1}^{N_{\mathrm{mocks}}}
\Big[P^{(k)}_{l;\;i}-\overline{P}_{l;\;i}\Big]
\Big[P^{(k)}_{l;\;j}-\overline{P}_{l;\;j}\Big]
\nonumber
\eeq
for the Nseries and QPM mocks. For BigMD, which only has one realization, we use the
covariance matrix of the Nseries realizations.  In the lower panel of Figure \ref{fig:NN_norm_resid}, 
we mark the monopole and quadrupole $k^\mathrm{NN}_{\chi^2}$ for the mock 
catalogs using the NN method. Arrows above the dashed line mark the monopole 
$k^\mathrm{NN}_{\chi^2}$ for Nseries (\nseriescolor), QPM (\qpmcolor) and 
BigMultiDark (\bmdcolor) catalogs. Similarly, the arrows below the dashed line 
mark the quadrupole $k^\mathrm{NN}_{\chi^2}$ for the mock catalogs. 
Averaged over the three mock catalogs, we get $k^\mathrm{NN}_{\chi^2} = 0.068 
\;\mathrm{and}\; 0.17\;h/\mathrm{Mpc}$ for the monopole and quadrupole respectively.

At $k = 0.2\;h/\mathrm{Mpc}$, the fiber collision residual for 
the monopole is over four times sample variance with average normalized 
residual of $4.4\%$ compared to the $0.9\%$ normalized sample variance. 
Moreover, we find that $k_{\chi^2} = 0.068\;h/\mathrm{Mpc}$, which is well below the maximum wavenumbers used typically in analyses. 
For the quadrupole, the fiber collision residual is approximately equivalent to 
sample variance at $k = 0.2\;h/\mathrm{Mpc}$ and $k_{\chi^2} = 0.17\;h/\mathrm{Mpc}$, but it quickly deteriorates with increasing $k$. 
Therefore for theoretical predictions that attempt to go beyond these scales, the effects of 
fiber collisions undoubtedly dominate the sample variance for both the power 
spectrum monopole and quadrupole and the NN method proves to be insufficient. In order to 
correct for this effect,  we next present our first approach: the `line-of-sight 
reconstruction' method.
\begin{figure*}
	\begin{center}
		\includegraphics[scale=0.575]{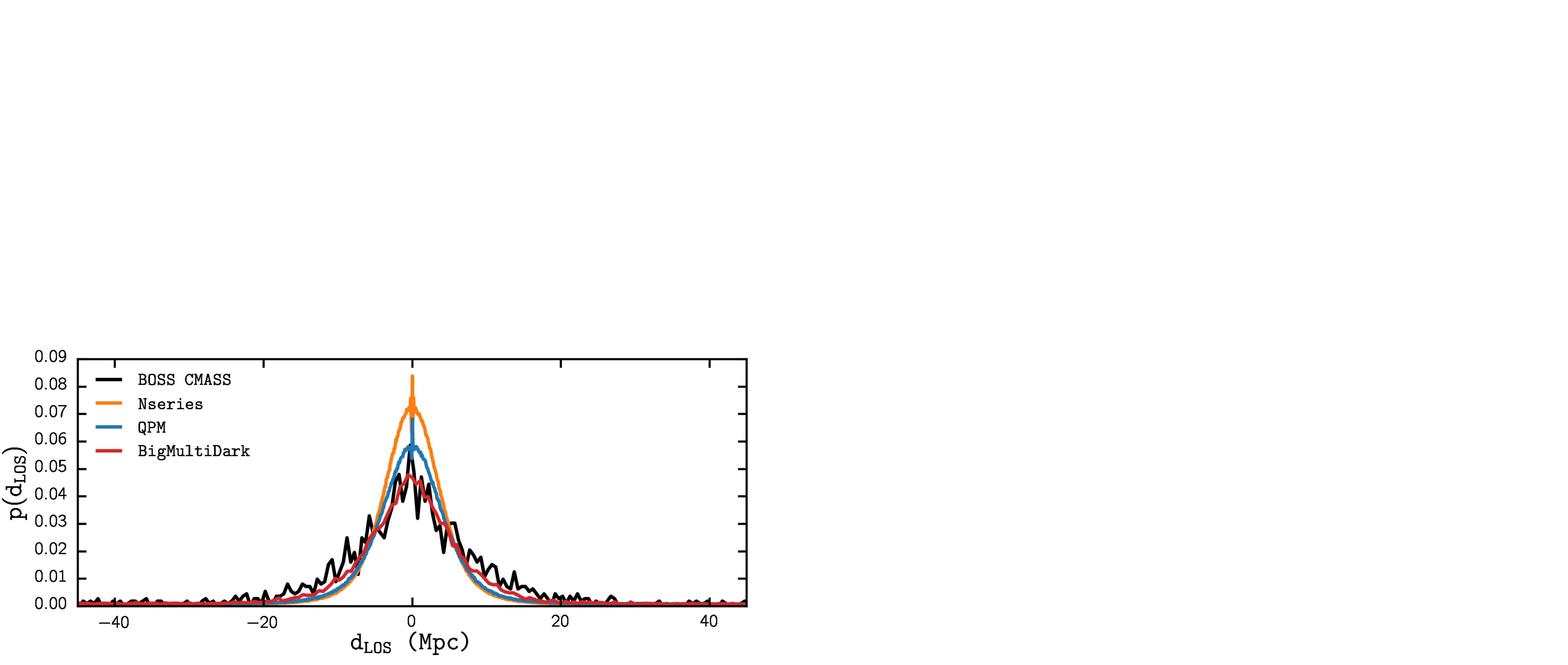}
			\caption{
            Normalized distribution of $d_{\mathrm{LOS}}$ for Nseries
            (\nseriescolor), QPM (\qpmcolor), and BigMultiDark (\tmcolor) 
            mock catalogs. The normalized $d_{\mathrm{LOS}}$ distribution of 
            BOSS DR12 is also plotted (\cmasscolor). The mock catalog distributions 
            have bin sizes of $\Delta d = 0.2\, \mathrm{Mpc}$, while the CMASS distribution has
            a bin size of $\Delta d = 0.5\, \mathrm{Mpc}$. The distribution extends beyond 
            the range of the above plot to $\sim \pm 500 \; \mathrm{Mpc}$. 
            In the discussion of Section \ref{sec:dlospeak}, we focus mainly 
            on the peak of the distribution at roughly
            $-20 \; \mathrm{Mpc} < d_\mathrm{LOS} < 20 \;\mathrm{Mpc}$.} 
            \label{fig:d_los}
	\end{center}
\end{figure*}
\subsection{Line-of-Sight Reconstruction Method} \label{sec:dlospeak}
\subsubsection{Line-of-Sight Displacement of Fiber Collided Pairs} \label{sec:dlos}
It is impossible to determine definitively from observed galaxy data whether 
individual fiber collided galaxies without resolved spectroscopic redshifts 
are correlated or chance alignments. However, the line-of-sight displacement of 
fiber collided galaxy pairs with resolved redshifts make it possible to model 
the overall impact fiber collisions have on displacing galaxies.

For the BOSS galaxy catalog, fiber collided pairs with resolved spectroscopic 
redshifts are mainly located in the overlapping regions (Section \ref{sec:catalog}).
For the simulated mock catalogs, fiber collisions are post-processed after 
the galaxy positions are generated. Therefore, 
all galaxies in fiber collided pairs have resolved redshifts. From these resolved 
redshifts we calculate the comoving line-of-sight displacement ($d_{\mathrm{LOS}}$) 
by taking the difference between the line-of-sight comoving distance of the 
resolved redshifts: 
\begin{equation}
d_{\mathrm{LOS}} = D_{\mathrm{C}} (z_1) - D_{\mathrm{C}} (z_2). 
\end{equation}
$D_{\mathrm{C}}(z)$ here is the line-of-sight comoving distance at $z$ 
(\citealt{Hogg:1999aa}), and $z_1$ and $z_2$ represent the resolved redshifts 
of the two galaxies in the fiber collided pair.

The normalized distributions of the calculated $d_\mathrm{LOS}$ for all
resolved fiber collided pairs are presented in Figure \ref{fig:d_los}
for Nseries (\nseriescolor), QPM (\qpmcolor), BigMultiDark (\bmdcolor), and BOSS DR12 
(\cmasscolor). The $d_{\mathrm{LOS}}$ distributions for all catalogs 
consist of two components: a peak roughly within the range $-20\;\mathrm{Mpc} 
< d_{\mathrm{LOS}} < 20\;\mathrm{Mpc}$ and a flat component (hereafter ``tail" component) 
outside the peak that extends to $d_{\mathrm{LOS}} \sim \pm 500 \;\mathrm{Mpc}$. The 
entire range of the distribution is not displayed in Figure \ref{fig:d_los}. 
For BOSS, as mentioned above, the $d_\mathrm{LOS}$ distribution only reflects the 
$d_\mathrm{LOS}$ values from galaxy pairs within the fiber collision angular scale 
with resolved spectroscopic redshifts, mostly from overlapping regions of the survey.

Galaxies within the same halo, due to their gravitational interactions at halo-scales, 
are more likely to be in close angular proximity with each other. These galaxies in 
over-dense regions cause the peak in the $d_{\mathrm{LOS}}$ distribution. The ``tail" 
component consists of chance aligned galaxy pairs that happen to be in close angular 
proximity in the sky. 

Focusing on the peak of the distribution, we note that it closely traces 
a Gaussian functional form. Therefore, we fit
\begin{equation} \label{eq:peak} 
	p(d_{\mathrm{LOS}}) = A \; e^{-{d_{\mathrm{LOS}}^2}/{2\sigma_\mathrm{LOS}^2}}
\end{equation}
for an analytic prescription of the $d_{\mathrm{LOS}}$ distribution peak as a 
function of $d_\mathrm{LOS}$ for each of the mock catalogs. We list the 
best-fit $\sigma_\mathrm{LOS}$ obtained by fitting Eq.~(\ref{eq:peak}) to the 
$d_{\mathrm{LOS}}$ distribution peak using 
MPFIT (\citealt{Markwardt:2009aa}) in Table \ref{tab:mpfit}. The parameter values
in Table \ref{tab:mpfit} and Figure \ref{fig:d_los} illustrate that the 
$d_{\mathrm{LOS}}$ distributions for the mock catalogs closely trace 
the BOSS DR12 distribution, which encourages our use of these mock 
catalogs in our investigation.  
\begin{table} 
  \caption{$d_{\mathrm{LOS}}$ Distribution Best-fit Parameters} \label{tab:mpfit}
  \begin{spacing}{1.5}
    \begin{center}
      \leavevmode
      \begin{tabular}{ccc} \hline \hline
      Catalog &$\sigma_\mathrm{LOS}$ ($\mathrm{Mpc}$) & $f_{\mathrm{peak}}$\\ \hline
      Nseries&3.88&0.69\\
      QPM&4.35&0.62\\
      BigMultiDark&5.47&0.60\\ 
      CMASS&6.56&0.70\\ \hline
      \end{tabular} \par
    \end{center}
  \end{spacing}
{\bf Notes}: Best-fit parameter $\sigma_\mathrm{LOS}$ (Eq.~\ref{eq:peak}) and peak fraction $f_{\mathrm{peak}}$ (Eq.~\ref{eq:fpeak}) for the $d_{\mathrm{LOS}}$ distributions in Figure \ref{fig:d_los}. 
\smallskip
\end{table}

Using the best-fit to the peak of the $d_{\mathrm{LOS}}$ distribution, 
we estimate the fraction of collided pairs that are within the peak as 
the ratio of pairs with $|d_\mathrm{LOS}| < 3\sigma_\mathrm{LOS}$ 
over the total number of pairs:  
\begin{equation} \label{eq:fpeak}
  f_{\mathrm{peak}} = \frac{\sum\limits_{|d_\mathrm{LOS}| < 3 \sigma_\mathrm{LOS}} p(d_{\mathrm{LOS}})}{N_{\mathrm{pairs}}}, 
\end{equation}
where $N_{\mathrm{pairs}}$ is the total number of fiber collided pairs. 
$f_\mathrm{peak}$ roughly corresponds to the fraction of galaxy pairs that are 
correlated. The $f_{\mathrm{peak}}$ values calculated for the mock catalogs are
listed in Table \ref{tab:mpfit}. They are consistent with the BOSS DR12 $f_\mathrm{peak}$. 

For the NN method of the previous section to be entirely correct, 
the $d_{\mathrm{LOS}}$ distribution in Figure \ref{fig:d_los} would 
have to be a delta function, which is clearly not the case. By simply 
incorporating the peak of the $d_{\mathrm{LOS}}$ distribution, we 
can significantly improve clustering statistics on small scales. Rather 
than placing the fiber collided galaxy on top of its nearest angular 
neighbor as the NN correction does, placing the fiber collided galaxy at a 
line-of-sight displacement, sampled from the peak of the $d_{\mathrm{LOS}}$
distribution, away from its nearest neighbor better reconstructs the 
galaxy clustering on small scales. 



Only $f_\mathrm{peak}$ of the 
collided pairs should be displaced, since only $f_\mathrm{peak}$ of the fiber 
collided pairs are correlated. Meanwhile, the other $(1-f_\mathrm{peak})$ pairs 
should retain their NN weights since they are uncorrelated. Displacing these galaxies as well according to the tail piece of the $d_{\mathrm{LOS}}$ distribution is not desirable because  in an object by object basis we do not know which galaxies should actually be in the tail of the distribution, thus we will be making large mistakes in $d_{\mathrm{LOS}}$ galaxy by galaxy. In addition, it is difficult to incorporate that these galaxies should be correlated with others and ignoring this modifies large-scale power. In our approach, the remaining $(1 - f_\mathrm{peak})$ fiber collided pairs are thus kept with their NN weights, and this is reflected in the shot noise correction of our estimator (Eq.~\ref{eq:roman_shotnoise}), which in turn makes connection to previous methods in the literature as we now discuss.



\subsubsection{Shot Noise Corrections} \label{sec:shotnoise} 
Measurements of the power spectrum are made on observations of discrete 
distributions of galaxies rather than continuous density fields. The 
discreteness contributes to the power spectrum. In order to correct for 
this contribution, galaxies are assumed to be Poisson samplings of the 
underlying distribution and a shot noise correction term 
is included in the power spectrum estimator \citep{Peebles:1980aa, Feldman:1994aa}. 

The expectation value of the shot noise term takes the following form \citep{Feldman:1994aa},
\begin{equation} \label{eq:integral_shotnoise}
  P_\mathrm{shot} = \frac{ (1+\alpha) \int d^3r \;\bar{n}({\bf r})w^2({\bf r})}{\int\limits^{ } d^3r \;\bar{n}^2({\bf r})w^2({\bf r})}. 
\end{equation}

\noindent Note that for the case of uniform weights ($w={\rm const.}$), 
constant number density and no random catalog this reduces to the standard 
shot-noise Poisson correction $P_\mathrm{shot} =\bar{n}^{-1}$. 
In practice the integrals in Eq.~(\ref{eq:integral_shotnoise}) can be 
written as discrete sums over the synthetic random catalog 
\citep{Feldman:1994aa}. $\int d^3r \; \bar{n}({\bf r}) ...$ 
is computed as $\alpha \sum_{\mathrm{ran}}...$. Then the shot noise term 
becomes, 
\begin{equation} \label{eq:fkp_shotnoise}
P^\mathrm{FKP}_\mathrm{shot} = \frac{(1+\alpha) \alpha \sum\limits_{\mathrm{random}} w_\mathrm{FKP}^2({\bf r})}{\alpha \sum\limits_{\mathrm{random}}^{ } \bar{n}({\bf r})\; w_\mathrm{FKP}^2({\bf r})}.
\end{equation} 
This however, represents the expectation value of the shot noise, not the actual value 
(\citealt{Hamilton:1997aa}) since all quantities involved are mean values (calculated through the random catalog). To use the full information provided by the data, the shot noise of the galaxies should be computed from the actual galaxy weights, not the randoms. This simply corresponds to taking the self-pairs in the power spectrum estimator, Eq.~(\ref{eq:roman_p0k}), which leads to Eq.~(\ref{eq:roman_shotnoise}) and we can rewrite here as,

\beqa \label{eq:ourshot}
P^\mathrm{Hahn+}_\mathrm{shot} &=& \frac{\sum\limits_{\mathrm{galaxy}} w^2_\mathrm{FKP}({\bf r})\, w^2_\mathrm{tot}({\bf r}) + \alpha^2 \sum\limits_{\mathrm{random}} w_\mathrm{FKP}^2({\bf r})}{\alpha \sum\limits^{ }_{\mathrm{random}} \bar{n}({\bf r})\, w_\mathrm{FKP}^2({\bf r})}\nonumber  \\ & & 
\eeqa
where $\alpha = (\sum_\mathrm{gal} w_\mathrm{tot} )/N_r$. 
We emphasize that this is {\em the} shot noise of the estimator. 
In other words, if one takes the limit $k \to \infty$, the 
estimator in Eq.~(\ref{eq:roman_p0k}) will approach this 
value if no shot-noise subtraction is applied. The systematic 
effects from completeness, redshift failures and fiber collisions are accounted for through $w_\mathrm{tot}$ of the observed galaxies. In our case, $w_\mathrm{tot}=w_\mathrm{sys}$ for the resolved $f_\mathrm{peak}$ fraction of galaxies that have been displaced away from their NN positions, while $w_\mathrm{tot}>w_\mathrm{sys}$ for the $(1-f_\mathrm{peak})$ fraction of galaxies that are deemed to be in the tail of the LOS distribution and are described by NN weights of the galaxies they collided with.


Recent work in the literature of power spectrum analysis modeled the effect of fiber collisions by solely modifying the shot noise term for the NN method \citep{Beutler:2014aa,Gil-Marin:2014aa}. This assumes that the effect of fiber collisions beyond NN weights is to alter the large-scale effective shot noise, and therefore that only the power spectrum monopole is affected since the quadrupole is free of shot noise. 
\cite{Beutler:2014aa} supplements the NN method with a shot noise correction term given by,
\begin{equation} \label{eq:florian}
P^\mathrm{B2014}_\mathrm{shot} = \frac{\sum\limits_{\mathrm{galaxy}} w^2_\mathrm{FKP}w_\mathrm{tot}({\bf r})w_\mathrm{sys}({\bf r}) + 
\alpha^2 \sum\limits_{\mathrm{random}} w_\mathrm{FKP}^2({\bf r})}
{\alpha \sum\limits_{\mathrm{random}}^{ } \bar{n} \; w_\mathrm{FKP}^2({\bf r})}.
\end{equation}
Note that in the first term of the numerator in this equation $w_\mathrm{fc}$ is only 
included in $w_\mathrm{tot}$ as it does not enter in $w_\mathrm{sys}$. 
It is also worth noting that \cite{Beutler:2014aa} ends up marginalizing over the value of the shot noise in their analysis, thus the impact of this particular choice is not straightforward.

Meanwhile, \cite{Gil-Marin:2014aa} constructs $P_\mathrm{shot}$ using two separate 
components: one for ``true pairs'' and the other for ``false pairs". 
The shot-noise contribution to the power from ``true pairs'' is the same as 
Eq.~(\ref{eq:florian}) while the ``false pairs'' shot-noise contribution is 
(same as Eq.~\ref{eq:ourshot}), 
\begin{equation} \label{eq:gm_falsepairs}
P^\mathrm{False}_\mathrm{shot} = \frac{\sum\limits_{\mathrm{galaxy}} w^2_\mathrm{FKP}w^2_\mathrm{tot}({\bf r}) + 
\alpha^2 \sum\limits_{\mathrm{random}} w_\mathrm{FKP}^2({\bf r})}
{\alpha \sum\limits^{ }_{\mathrm{random}} \bar{n} \; w_\mathrm{FKP}^2({\bf r})}.
\end{equation}
\cite{Gil-Marin:2014aa} calculates the total $P_\mathrm{shot}$ as the 
weighted combination of $P^\mathrm{True}_\mathrm{shot}$ and
$P^\mathrm{False}_\mathrm{shot}$: 
\begin{equation} \label{eq:gm_shot}
P^\mathrm{GM2014}_\mathrm{shot} = (1- x_\mathrm{PS}) P^\mathrm{True}_\mathrm{shot} +
x_\mathrm{PS}\, P^\mathrm{False}_\mathrm{shot}
\end{equation}
In their analysis, \cite{Gil-Marin:2014aa} use $x_\mathrm{PS} = 0.58$, 
which they infer by measuring the difference between 
the true and the fiber-collided power spectrum monopole in the $\mathtt{PTHalos}$ 
galaxy mock catalogs (\citealt{Manera:2013aa}). Unfortunately, since the true 
power spectrum is the measurement we are trying to recover from the observations, the 
$x_\mathrm{PS}$ parameter cannot be inferred or validated from the actual 
BOSS observations. Moreover, one might worry about relying $\mathtt{PTHalos}$ or similar methods that are not based on 
high resolution N-body simulations, to extract corrections for fiber collisions that depend on small-scale power. An extension of this approach is used in recent BOSS analyses  \citep{Beutler:2016aa,Grieb:2016aa,Gil-Marin:2016aa} where Eq.~(\ref{eq:gm_shot}) is used and is supplemented with a marginalization over the shot noise value. However, as we discussed above, this has no effect in the quadrupole power spectrum, which remains the same as in the NN method.

At this point it is worth casting our ``line-of-sight reconstruction" (LRec) 
method in similar language to the methods we just discussed. We treat the 
``true pairs" (what we called peak-pairs) by displacing them according to 
the peak LOS distribution, which modifies all the power spectrum multipoles, 
and use the NN method for the ``false pairs" (pairs in the tail of the LOS 
distribution). Our shot noise correction is not adjusted, rather it is the 
true shot noise from the estimator. 
We now discuss the implementation and performance of our  LRec fiber collision method.

\begin{figure*}
\begin{center}
\includegraphics[scale=0.55]{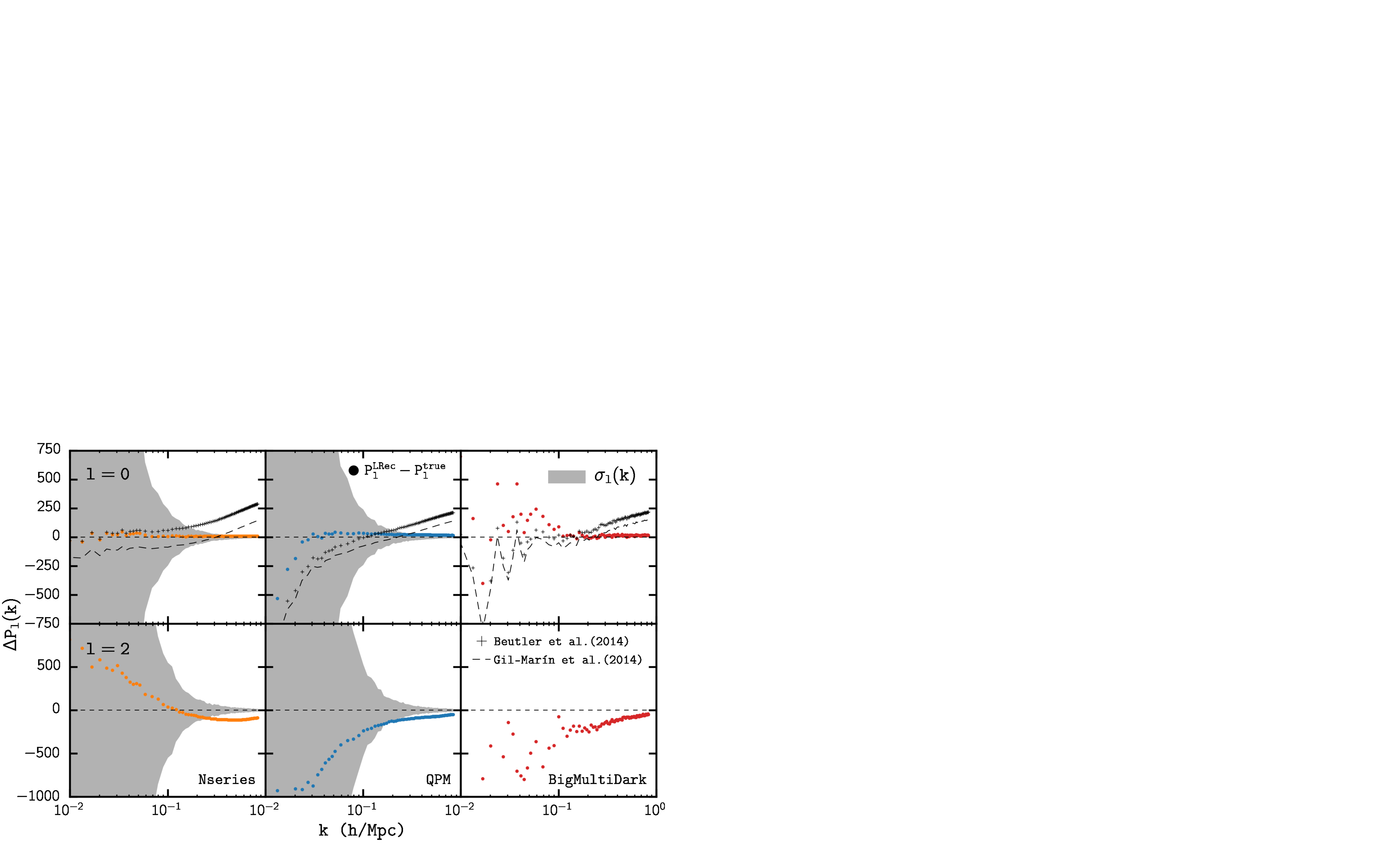} 
\caption{The power spectrum residual of the line-of-sight reconstruction (LRec) 
method (Section \ref{sec:dlospeak}), 
$\Delta P_\ell \equiv P_l^\mathrm{LRec} -P_l^\mathrm{true}$, 
for the monopole (top) and quadrupole (bottom) power spectra of the 
Nseries (left), QPM (middle), and BigMultiDark (right) mock catalogs. 
We again plot the Nseries and QPM sample variances, $\sigma_l(k)$.
The residuals for the monopole show good agreement between 
$P_0^\mathrm{LRec}$ and $P_0^\mathrm{true}$ for the entire $k$ range. 
For the quadrupole, while the LOS Reconstruction method improves the residuals 
compared to the NN method at small scales ($k > 0.2\;h/\mathrm{Mpc}$), 
the residuals remain comparable to sample variance at $k=0.2\;h/\mathrm{Mpc}$.
In the top panels, we include the residuals from the fiber collision 
correction methods of \cite{Beutler:2014aa} (plus) and \cite{Gil-Marin:2014aa} 
(dashed). Both these corrections supplement the NN method with adjustments to 
the constant shot noise term of the estimator. As a result, they fail to correct for the
$k$ dependence of the effect and are insufficient in accounting for fiber collisions at small scales. 
\cite{Beutler:2014aa} marginalizes over the correction in their analysis, 
so we plot the correction from Eq.~(\ref{eq:florian}) offset by $-250$ to match 
low-$k$ residuals on the left panel.} 
\label{fig:peaksn}
\end{center}
\end{figure*}

\begin{figure}
\begin{center}
\includegraphics[scale=0.4]{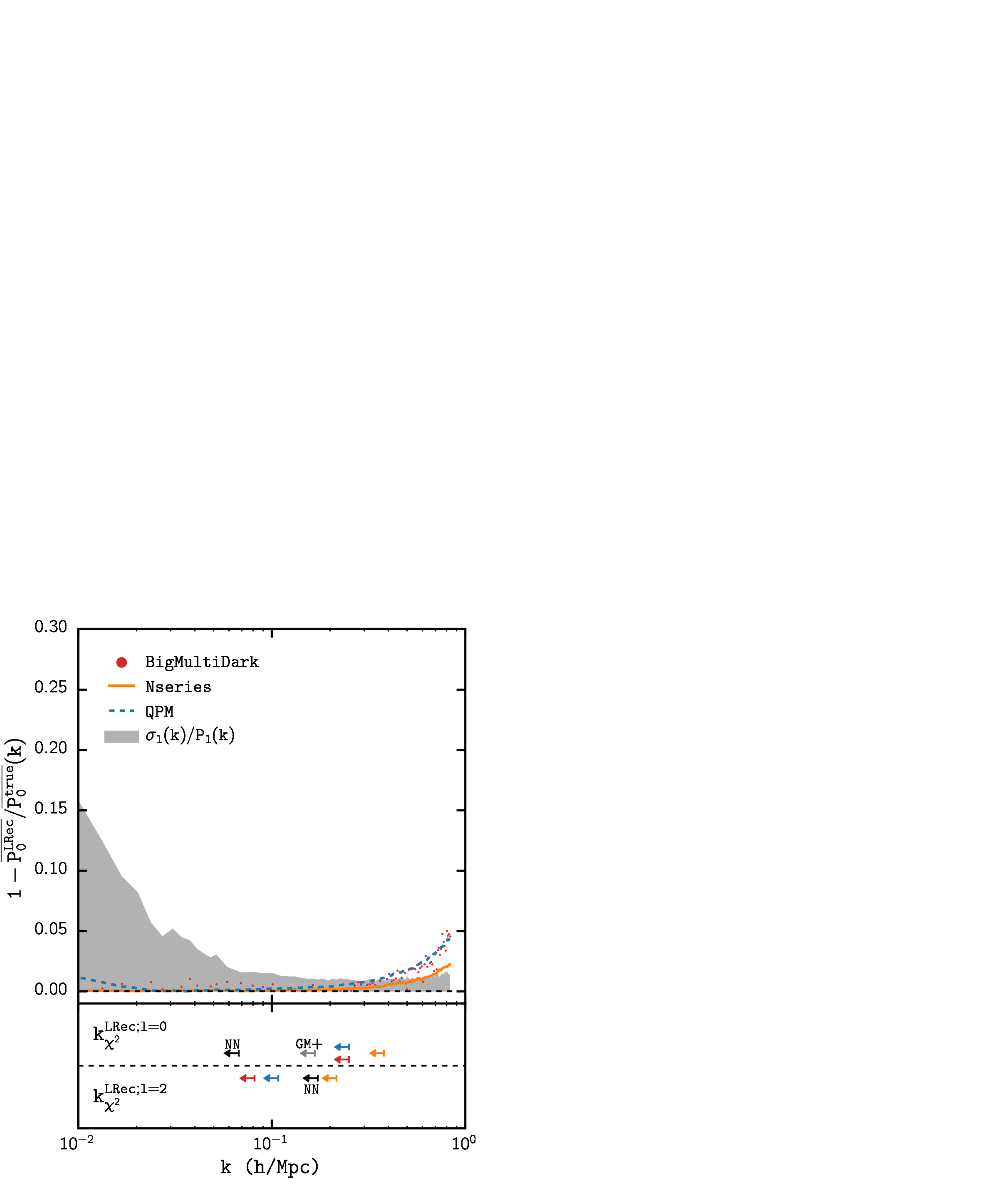} 
\caption{{\it Top Panel}: The normalized residual, 
$1 - P_l^\mathrm{LRec}/P_l^\mathrm{true}$, 
for the Nseries (\nseriescolor), QPM (\qpmcolor), and BigMultiDark (\bmdcolor)
monopole power spectra. The normalized sample variance $\sigma_l / P_l(k)$ 
(gray shaded region) of the Nseries mocks is plotted for comparison. At 
$k = 0.1 \;h/\mathrm{Mpc}$, where the NN method residuals exceeds sample 
variance, the average normalized residual for the LRec method is $0.25\%$ 
compared to $1.5\%$ normalized sample variance. In fact, the average 
residual stays below the sample variance until $k = 0.53\;h/\mathrm{Mpc}$.\\
{\it Bottom Panel}: We mark $k^\mathrm{LRec}_{\chi^2}$ for the monopole (arrows above the 
dashed line) and quadrupole (arrows below the dashed line). 
The average $k^\mathrm{LRec}_{\chi^2}$ for the mock catalogs
are $0.29$ and $0.14\;h/\mathrm{Mpc}$ for the monopole and quadrupole respectively. 
For comparison, we mark $k^\mathrm{NN}_{\chi^2}$ (black) from Section \ref{sec:fc_pk}. 
We also include $k_{\chi^2}$ of the \cite{Gil-Marin:2014aa} correction method (gray) for 
the monopole. The LOS reconstruction method significantly 
extends $k_{\chi^2}$ beyond that of the NN method and \cite{Gil-Marin:2014aa} for $l=0$. 
However, it does not improve $k_{\chi^2}$ for the quadrupole. 
} 
\label{fig:dlospeak_norm_resid}
\end{center}
\end{figure}
\subsubsection{In Practice}
We first begin with fiber collided mock catalogs with the NN 
fiber collision weights that accurately simulate the effects of 
fiber collisions on the actual BOSS observations. From this catalog, we
construct the $d_\mathrm{LOS}$ distribution, as described in Section 
\ref{sec:dlos} and fit for the best-fit parameters $\sigma_\mathrm{LOS}$ 
and $f_\mathrm{peak}$ of Eq.~(\ref{eq:peak}).

We select $f_\mathrm{peak}$ of the fiber collided galaxy pairs in the catalog 
and designate them as correlated pairs that lie within the peak of the 
$d_\mathrm{LOS}$ distribution. We refer to these fiber collided pairs as 
``peak-assigned". At this point, each of these pairs, based on their NN 
weights, consist of the ``nearest-neighbor" galaxy with $w_\mathrm{fc} > 1$ 
and the ``collided" galaxy with $w_\mathrm{fc} = 0$. We discard the collided galaxy
since the redshifts of collided galaxies are not known in actual observations. 

Next for each of the nearest-neighbor galaxies in peak-assigned pairs, we 
place a new galaxy with $w_\mathrm{fc} = 1$ at a displacement $d_\mathrm{peak}$ 
away from it along the line-of-sight but at the same angular position. The $d_\mathrm{peak}$ 
value is sampled from a Gaussian with best-fit $\sigma_\mathrm{LOS}$ from 
Table \ref{tab:mpfit}. The $w_\mathrm{fc}$ of the ``nearest-neighbor" galaxy 
is then reduced by $1$. This process is repeated, in the cases of triplets 
or higher with $w_{\rm fc} > 2$, until all the nearest-neighbor galaxy in 
peak-assigned pairs have $w_\mathrm{fc}=1$. The resulting {\em total} catalog will have fewer 
galaxies with $w_\mathrm{fc} > 1$ compared to the initial fiber collided catalog.
However, the total statistical weight ($\sum_\mathrm{gal} w_\mathrm{tot}$) of 
the catalog, being equal to the total number of galaxies before the collisions
are applied, is conserved. 

Now that we have the ``LOS reconstructed" mock catalog, we measure its power 
spectrum monopole and quadrupole ($P^\mathrm{LRec}_l$). In Figure 
\ref{fig:peaksn} we present the power spectrum residual, 
$(P^\mathrm{LRec}_l-P^\mathrm{true}_l)$,  
for $l = 0$ and $2$ of the LOS Reconstruction method power spectrum averaged over all the available realizations. 
We again include the Nseries and QPM sample variance, $\sigma_l(k)$ 
(grey shaded region) for comparison.  In Figure \ref{fig:dlospeak_norm_resid}, 
we normalize both the residuals and the sample variance by $P_0^\mathrm{true}$ 
to better compare $P_0^\mathrm{LRec}$ and $P_0^\mathrm{true}$ at different scales 
and to highlight the small scales. 

For the monopole, at the scale where $P^\mathrm{NN}_0$ deviates from 
$P^\mathrm{true}_0$ by more than the sample variance ($k \sim 0.1\;h/\mathrm{Mpc}$), 
Figure \ref{fig:peaksn} shows that the LOS reconstructed residual is well within the 
sample variance, $P^\mathrm{LRec}_0 - P^\mathrm{true}_0 < 0.17\, \sigma_0$. 
Even at the smallest scales measured for our monopole measurements 
($k = 0.83\;h/\mathrm{Mpc}$), well beyond the scales that can be predicted from current models based on perturbation theory, the normalized residuals for 
the LOS reconstructed method remains at $3.7\%$. At $k \sim 0.2\;h/\mathrm{Mpc}$, the average normalized 
residual is $0.19\%$ compared to the $0.9\%$ normalized sample variance. 
When we calculate the $k_{\chi^2}$ of the LOS reconstruction method for 
the three mock catalogs, as we did for the NN method in Section \ref{sec:fc_pk}, 
we get the average $k_{\chi^2}^\mathrm{LRec} = 0.29\;h/\mathrm{Mpc}$ for the monopole. 
For each of the mocks, we mark $k_{\chi^2}^{\mathrm{LRec};\;l=0}$ in the lower panel 
of Figure \ref{fig:dlospeak_norm_resid} above the dashed horizontal line. 

For the monopole, we also include the residuals from the 
fiber collision correction methods of \cite{Beutler:2014aa} (pluses) and 
\cite{Gil-Marin:2014aa} (dashed) in Figure \ref{fig:peaksn}. 
Both these analyses correct for fiber collisions by adjusting the constant 
shot noise term in the estimator in addition to the NN method (Section \ref{sec:shotnoise}).
However, as the NN method power spectrum residuals reveal in Figure \ref{fig:fc_pk},
the effect has a $k$ dependence, especially at $k > 0.1 \;h/\mathrm{Mpc}$.  
So while these corrections can reduce the residuals to within sample variance 
at large scales, they fail to account for the $k$ dependence, 
which quickly goes on to dominate sample variance at smaller scales, 
$k > 0.1 \;h/\mathrm{Mpc}$.  

We note that instead of using a fixed value for the constant shot noise
as \cite{Gil-Marin:2014aa} does, \cite{Beutler:2014aa} marginalize over 
the constant term in their analysis. To  reflect this,
we offset the power spectrum residual we get using Eq.~(\ref{eq:florian}) 
by $-250$ in Figure \ref{fig:peaksn} to force agreement at $k\to 0$ 
in the Nseries case. For simplicity, we only calculate $k_{\chi^2}$ 
for the \cite{Gil-Marin:2014aa} correction method using the mock catalogs: 
$k_{\chi^2}^\mathrm{GM+} = 0.17 \;h/\mathrm{Mpc}$ (gray arrow; Figure \ref{fig:dlospeak_norm_resid}), 
which is significantly lower than that of the LOS Reconstruction method. 
Compared to either method, the LOS reconstruction 
method better accounts for fiber collisions at all scales. 
Furthermore, as already discussed the methods of \cite{Beutler:2014aa} 
and \cite{Gil-Marin:2014aa} do not provide corrections for the power 
spectrum quadrupole or higher multipoles, thus Figure~\ref{fig:fc_pk} 
still applies for $\ell=2$.   

From Figure~\ref{fig:peaksn} we see that for the quadrupole, the LOS reconstruction method does not 
sufficiently improve corrections for fiber collisions compared to the NN method.
The residuals for $k > 0.2\;h/\mathrm{Mpc}$ are improved compared to Figure~\ref{fig:fc_pk}; however, they 
still exceed the sample variance. Unfortunately, these improvements on small scales
come at the cost of increased residuals on large scales. In the $k_{\chi^2}$ marked 
in Figure \ref{fig:dlospeak_norm_resid} (below the dashed line), we see that the
increased residuals at large scales actually make the average $k_{\chi^2}^\mathrm{NN} > 
k_{\chi^2}^\mathrm{LRec} = 0.14\;h/\mathrm{Mpc}$ for the quadrupole, although there is significant dispersion between the different simulations with Nseries showing improvements when compared to the NN method while the other two showing worse performance. Consequently, neither the LOS reconstruction method nor the NN method sufficiently 
account for fiber collisions in the power spectrum quadrupole. 

The shortcomings of the LOS reconstruction method for the quadrupole compared 
to the monopole does not come as a surprise since the quadrupole is more 
sensitive to getting the correct LOS displacements galaxy by galaxy 
(not just statistically), as these modify the fingers-of-god effect. 
In order to make further progress with this method one would have to 
determine for each galaxy the most likely halo in which it lives 
(this could be nearby or a distant, chance alignment), determine its 
velocity dispersion and then assign a LOS displacement consistent 
with the dispersion and the observed LOS distribution. 

Let us now discuss a few attempts that we have implemented along these lines. 
The first is incorporating more information about the fiber collided pairs 
in order to better classify 
correlated and chance alignment pairs. For example, information about 
larger scale galaxy environment in the form of the $N^{th}$ nearest neighbor 
distance ($d_{nNN}$), can be included to parameterize the $\sigma_\mathrm{LOS}$ 
and $f_\mathrm{peak}$ (Table \ref{tab:mpfit}) as a function of $d_{nNN}$. 
The $d_{nNN}$ in this case is the distance of the $n^{th}$ nearest 
neighbor of the nearest-neighbor galaxy within the fiber collided pair.  
Another way the LOS reconstructed method can be improved is by utilizing 
the photometric redshifts of the collided galaxies to improve the 
correlated/change alignment pair classification. 

We explored the LOS reconstructed method with both of these improvements 
on the mock catalogs. We find that there is indeed a significant correlation 
between $d_{nNN}$ and the parameters $\sigma_\mathrm{LOS}$ and 
$f_\mathrm{peak}$, which can be exploited. Also, photometric redshifts assigned 
to collided galaxies based on the 
$|z_\mathrm{spec} - z_\mathrm{photo}|/(1 + z_\mathrm{spec})$
of actual BOSS photometric redshift catalogs improves classification of 
correlated versus chance alignment fiber collided pairs, as well. These 
improvements bring the normalized residuals of the monopole to $\sim 1\%$
at $k = 0.83\;h/\mathrm{Mpc}$. However, the improvement in the fiber collision 
correction for the quadrupole is marginal; the effect of fiber collisions 
at $k = 0.2\;h/\mathrm{Mpc}$ is still comparable to the sample variance. So 
even with these improvements the LOS reconstructed method is insufficient. 

Furthermore, for the Nseries mocks, we find that if we use the LOS reconstructed method with 
perfectly classified correlated and chance alignment pairs, the 
residual is roughly half the sample variance at $k \sim 0.2\;h/\mathrm{Mpc}$
and greater than sample variance at $k > 0.35\;h/\mathrm{Mpc}$.
The displacement of the collided galaxy by $d_\mathrm{LOS}$ sampled from Eq.~(\ref{eq:peak}) 
alone causes the power spectrum quadrupole to deviate from the true value
at small scales. A method such as the LOS reconstructed method for the quadrupole 
would require more sophisticated modeling of the fiber collided galaxy pairs that capture the displacements in an object by object basis. 

As a result of the shortcomings of the LOS reconstructed method for the power spectrum 
quadrupole, we now present a complementary approach in dealing with fiber collision 
in power spectrum multipole analyses, which rather than attempting to correct the 
data before making measurements, computes theoretical predictions of the fiber-collided 
power spectrum multipoles. 

\begin{figure*}
\begin{center}
\includegraphics[scale=0.3]{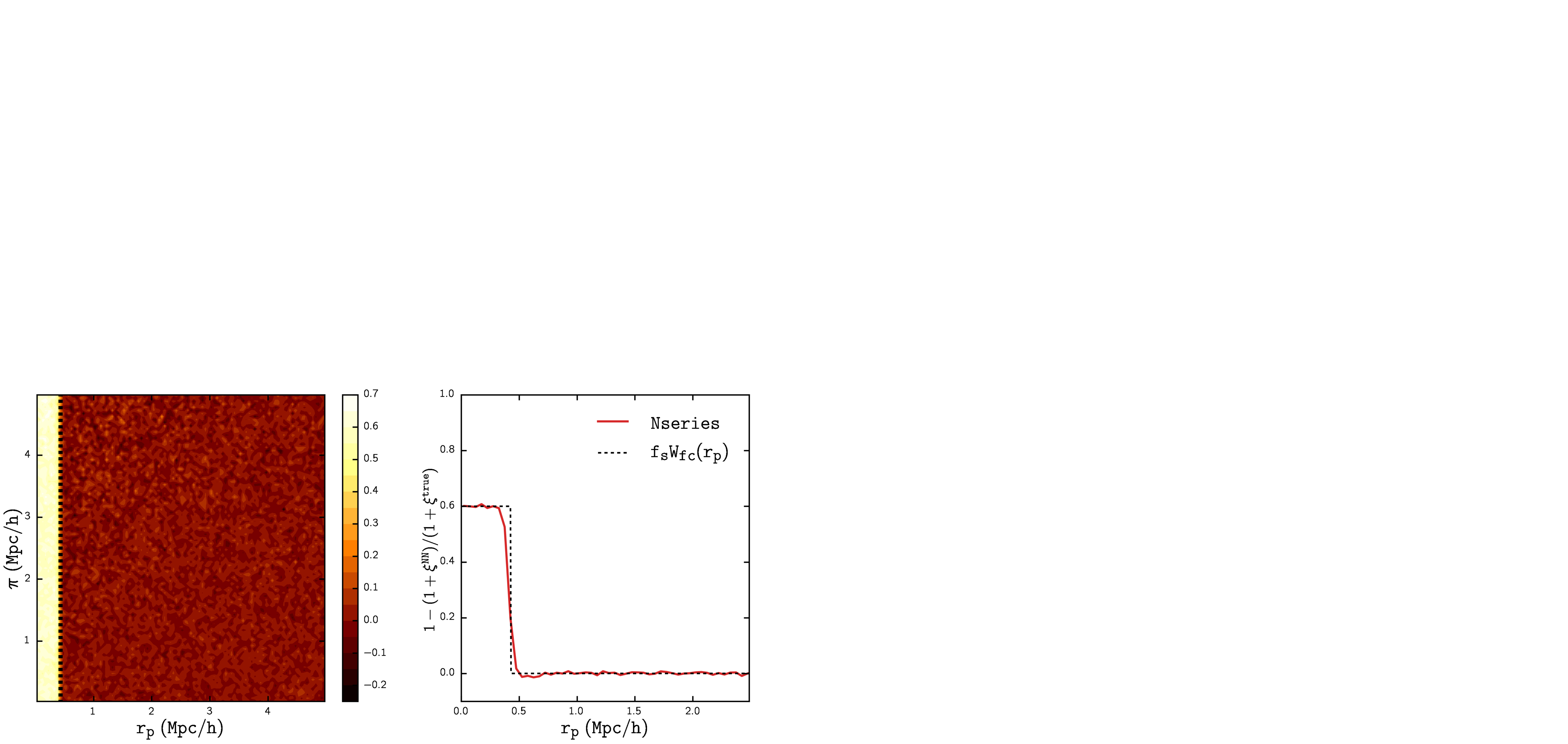} 
\caption{$1 - (1 + \xi^\mathrm{NN})/(1+\xi^\mathrm{true})$ as a 
function of transverse displacement, $r_p$, and line-of-sight 
displacement $\pi$ (left). The color bar represents the value of this quantity. Note there is no detectable dependence on $\pi$.
The dashed vertical line (black) represents the constant 
$r_p = D_\mathrm{fc}(z=0.55)$ (Section \ref{sec:fourier}). 
We also plot $1 - (1 + \xi^\mathrm{NN})/(1+\xi^\mathrm{true})$
projected along $\pi$ (right). In the left panel, the $r_p = D_\mathrm{fc}(z=0.55)$ 
vertical line and the sharp cut-off of the contour show good agreement with the expected characteristic scale. 
In the right panel, the projected $1 - (1 + \xi^\mathrm{NN})/(1+\xi^\mathrm{true})$
is in good agreement with $f_s W_\mathrm{fc}(r_p)$. The agreement in both panels
justify the characterization of the effect of fiber collisions on the 2PCF in 
Eq.~(\ref{eq:tophat_2pcf}).}
\label{fig:2pcf_tophat}
\end{center}
\end{figure*}
\subsection{Effective Window Method} \label{sec:fourier}
The LOS Reconstruction method corrects for fiber collisions in the observed galaxy positions 
 in order to estimate the systematics-free true power 
spectrum. In power spectrum analyses, this true power spectrum estimate
can be compared to model power spectrum for cosmological parameter inference. 
Alternatively, however, the observed fiber collided power spectrum can be 
compared to the model power spectrum with the effect of fiber collisions imposed 
on it. This is the approach we follow from now on.

We proceed as follows. In Section \ref{sec:tophat_theory} we find that the effect of fiber collisions 
on the two-point correlation function can be well approximated by a simple 
analytic expression. Using this, we accurately estimate the 
effect of fiber collisions on the power spectrum in Fourier space. The effect 
is a function of the true power spectrum and depends significantly on the 
power spectrum at small scales, which cannot reliably be modeled from first principles. As a result, in Section~\ref{sec:tophat_practice}, we 
present a practical approach to circumvent this issue and account for the effect of fiber collisions in 
power spectrum analyses. 

\subsubsection{In Theory} \label{sec:tophat_theory}
In the BOSS galaxy catalog, which spans the redshifts $0.43 < z < 0.7$, 
the comoving distance of the $62 \arcsec$ fiber collision angular scale 
($D_\mathrm{fc}$) ranges from $0.35\;\mathrm{Mpc}$ to $0.52\;\mathrm{Mpc}$. 
Given the relatively small variation in $D_\mathrm{fc}$, we assume 
that throughout the survey redshift the physical scale remains
constant as $D_\mathrm{fc}(z \sim 0.55) = 0.43 \mathrm{Mpc}$, at the median 
redshift of the survey. If the physical scale of fiber collisions is constant, 
fiber collisions will affect the two-dimensional configuration space two-point 
correlation function, $\xi(r_p, \pi)$, through its effect on galaxy pairs with 
transverse separations $r_p < D_\mathrm{fc}$. As no pairs will be found 
below this characteristic scale, $\xi(r_p, \pi)$ will be -1 
for $r_p < D_\mathrm{fc}$, and note that  the same is true for the two-point function in the NN method (since small-$r_p$ pairs are collapsed into zero separation described by weights). On the other hand, at large scales we can approximate $\xi(r_p, \pi)$ by the NN method which preserves the large-scale angular correlation function, thus 
the effect of fiber collisions on $\xi(r_p, \pi)$ can be analytically characterized by  the following relation between the true and the NN two-point functions,
\begin{equation} \label{eq:tophat_2pcf}
\frac{1 + \xi^\mathrm{NN}(r_p, \pi)}{1 + \xi^\mathrm{true}(r_p, \pi)} =1 -  f_s W_\mathrm{fc}(r_p)
\end{equation}
where $W_\mathrm{fc}(r_p)$ represents the top-hat function
\begin{spacing}{1.5}
\begin{equation} \label{eq:tophat}
W_\mathrm{fc}(r_p) = 
\begin{cases}
1 & \text{if}\ r_p < D_\mathrm{fc} \\
0 & \text{otherwise}
\end{cases}
\end{equation}
\end{spacing}
\noindent and $f_s$ represents the fraction of the survey area affected by fiber 
collisions. Note in Eq.~(\ref{eq:tophat_2pcf}) we have assumed that we can linearly superpose the contributions to the two-point function from regions with and without collisions, and a key property of Eq.~(\ref{eq:tophat_2pcf}) is that its right hand side does not depend on $\pi$, something we test explicitly below. 
In the BOSS, $f_s$ is precisely known because 
it corresponds to the fraction of the survey geometry that suffers from 
fiber collisions. These are the regions that do not have overlapped tiling 
(Section \ref{sec:catalog}). For BOSS DR12 $f_s = 0.6$. 

We measure $\xi^\mathrm{NN}$ and $\xi^\mathrm{true}$ for the Nseries 
mock catalogs using the $\mathtt{CUTE}$ software (\citealt{Alonso:2012aa}), which 
uses the standard \cite{Landy:1993aa} estimator. $\xi^\mathrm{NN}$ 
is calculated from the NN fiber collided Nseries mocks while 
$\xi^\mathrm{true}$ is calculated from the Nseries mocks without fiber 
collisions. Using the measured $\xi^\mathrm{NN}$ and $\xi^\mathrm{true}$, 
we plot 
$1- (1+\xi^\mathrm{NN})/(1 + \xi^\mathrm{true})$ averaged over realizations as 
a function of $r_p$ and $\pi$ (left) and its projection 
along $\pi$ (right) in Figure \ref{fig:2pcf_tophat}. The dashed vertical 
line (black; left) marking $r_p = D_\mathrm{fs}(z=0.55)$ and $f_s W_\mathrm{fc}(r_p)$ 
(black dashed; right) are plotted for comparison. The agreement 
between the $\xi(r_p, \pi)$ contours and the $r_p = D_\mathrm{fc}(z=0.55)$ 
cutoff along with the agreement between the projection and 
$f_s W_\mathrm{fc}(r_p)$ justify our assumption of a constant physical 
fiber collision scale. The exact survey tiling of the BOSS sample is 
imposed on the Nseries mocks, so we expect Figure \ref{fig:2pcf_tophat} 
to hold for the BOSS observations. The left panel illustrates 
the $\pi$-independence of the left hand side of Eq.~(\ref{eq:tophat_2pcf}). 
The right panel demonstrates that 
$1- (1+{\xi^\mathrm{NN}})/(1 + {\xi^\mathrm{true}})$ projected
along $\pi$ agrees remarkably well with a top-hat function. 

In principle, however,
$W_\mathrm{fc}$ is not necessarily a top-hat function. In fact,
in eBOSS, due to the complex targeting scheme involving ``knock-outs'' from 
higher priority targeting samples, $W_\mathrm{fc}$ will not be top-hat function 
(Zhai et al. in prep). However, these complications are not present in our implementation of collisions; the reason for the deviations from a top-hat function here can be thought as arising from a sum of top-hats of slightly different radii along the line of sight (for fixed angular scale) weighted by the probability of collisions at each depth, leading to a smoother transition than a sharp top-hat function. In principle, our formalism  can be improved by including this numerical profile rather than a top-hat, as we shall mention below (see discussion after Eq.~\ref{DeltaPell2}).

With the confirmation of Eq.~(\ref{eq:tophat_2pcf}), we solve for $\xi^\mathrm{NN}$   
:
\beqa
\xi^\mathrm{NN}(r_p, \pi) &=& \xi^\mathrm{true}(r_p, \pi) - f_s W_\mathrm{fc}(r_p)\ (1 + \xi^\mathrm{true}(r_p, \pi)),
\nonumber \\ & & 
\eeqa
and to get an expression for the power spectrum, we Fourier transform to get 
\beqa 
\label{eq:tophat_pk}
\Delta P({\bf k}) &\equiv& P^\mathrm{NN}({\bf k}) - P^\mathrm{true}({\bf k})  \nonumber \\ & & 
= - f_s\, {W_\mathrm{fc}}({\bf k})- f_s \int {\mathrm{d}^3q\over (2\pi)^3} P({\bf q})\, {W_\mathrm{fc}}({\bf k} - {\bf q}). \nonumber \\ & & 
\eeqa
We see that the effect of fiber collisions on the true power spectrum 
can be characterized by two terms: Fourier transform of the 
top-hat function (corresponding to chance collisions) 
and the power spectrum convolved with the top-hat function (corresponding to 
physically correlated pairs). We refer to these two terms as $\Delta P^\mathrm{uncorr}$ and 
$\Delta P^\mathrm{corr}$ respectively. Note that none of these terms is independent of $k$.

\begin{figure*}
\begin{center}
\includegraphics[scale=0.5]{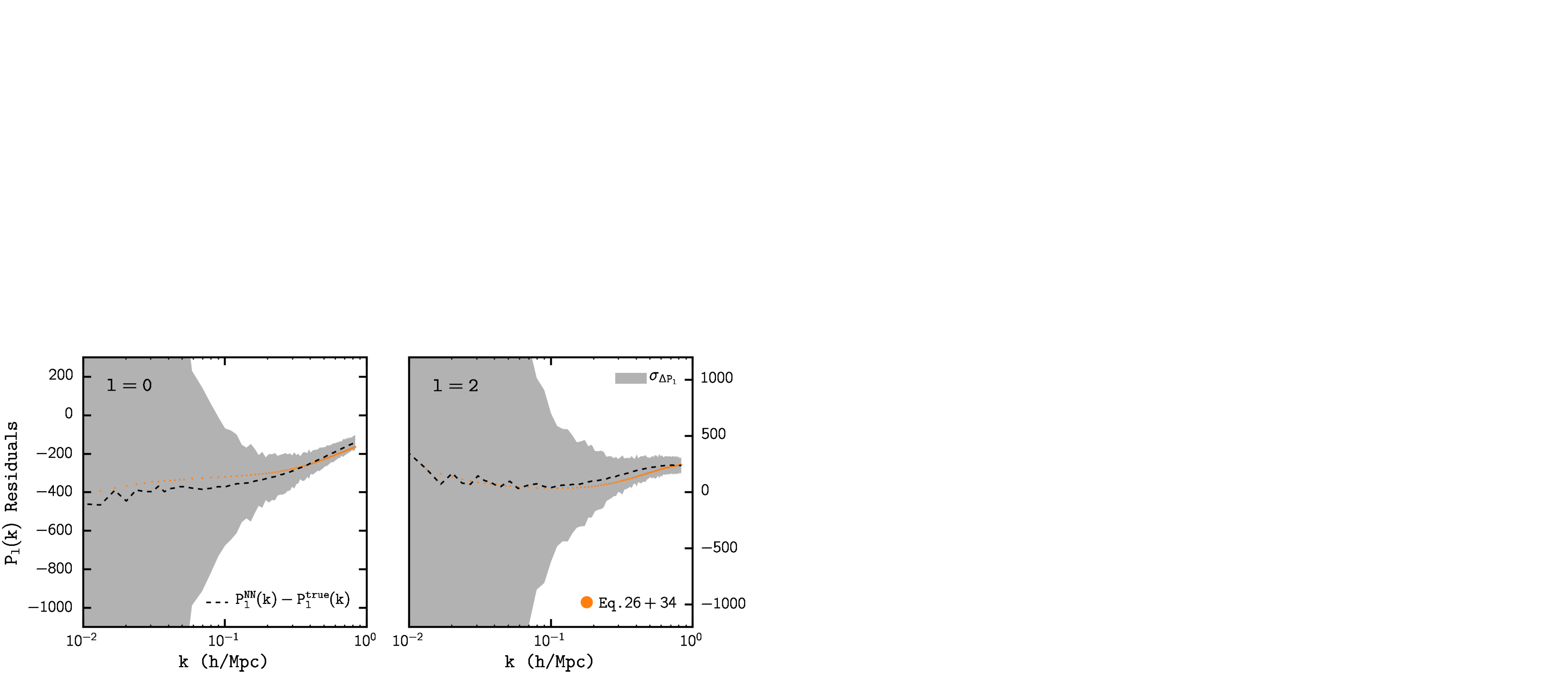}
\caption{Comparison of the power spectrum residuals from NN-corrected fiber collisions 
$\Delta P_l = P_l^\mathrm{NN} - P_l^\mathrm{true}$ (dashed black) with 
the $\Delta P_l$ from the effective window method obtained by adding Eqs.~(\ref{DeltaPellUC}) and~(\ref{DeltaPell2}) (\nseriescolor) for the monopole (left)
and quadrupole (right). The standard deviation of the power spectrum residual, 
$\sigma_{\Delta P_l}$, for the Nseries mock catalogs is shaded in gray.}
\label{fig:delP}
\end{center}
\end{figure*}

The first term, $\Delta P^\mathrm{uncorr}$, can be easily obtained:
\begin{align} \label{eq:delp_uncorr}
\Delta P^\mathrm{uncorr} &=  - f_s\, \widehat{W_\mathrm{fc}}({\bf k}) = 
-f_s \int e^{i {\bf k} \cdot {\bf r}} \: W_\mathrm{fc}({\bf r}) \, d^3{r} \nonumber \\
& = -f_s \ 2 \pi \delta_D(k_\parallel) \; \pi D_\mathrm{fc}^2 \; W_\mathrm{2D}(k_\perp D_\mathrm{fc}). 
\end{align}
where $W_\mathrm{2D}(x) \equiv 2 J_1(x)/x$ is the top-hat function in 2D (a cylinder), and 
 $J_1$ is a Bessel function of the first kind and of order $1$. The multipole 
contributions of Eq.~(\ref{eq:delp_uncorr}) are then 
\beqa
\Delta P^\mathrm{uncorr}_l(k) & = & -f_s \  (2l+1) \mathcal{L}_l(0) \,  {(\pi D_\mathrm{fc})^2 \over k} \;  W_\mathrm{2D}(k D_\mathrm{fc}),
\nonumber \\ & & 
\label{DeltaPellUC}
\eeqa
where ${\cal L}_l$ are the Legendre polynomials. The $k^{-1}$ prefactor here, arising from the delta function in Eq.~(\ref{eq:delp_uncorr}) is an approximation for scales smaller than the survey size, since the delta function follows from assuming we can integrate up to infinity along the line of sight in Eq.~(\ref{eq:delp_uncorr}).
Equation~(\ref{DeltaPellUC}) gives a correction that alternates in sign as a function of multipole $l$.
Note that since for practical purposes $k D_\mathrm{fc} \ll 1$, we can expand 
\beqa
\Delta P^\mathrm{uncorr}_l(k)  &= & -f_s \pi D_\mathrm{fc}^2  \,  \Big({2\pi \over k}\Big) \ {(2l+1) \over 2}\, \mathcal{L}_l(0) \nonumber \\
& & \times \Big(1 - \frac{(k D_\mathrm{fc})^2}{8} + \ldots \Big),
\label{DeltaPuncorrexp}
\eeqa
and for scales involved in typical analysis the first term suffices, which means that the uncorrelated piece of fiber collisions decays as $k^{-1}$ across the relevant range of scales. The magnitude of this uncorrelated  effect (chance collisions) is small, given by the effective survey area affected by fiber collisions $f_s \pi D_\mathrm{fc}^2$ times the wavelength of perturbations $2\pi/k$.

For the correlated piece $\Delta P^\mathrm{corr}$, we see from Eqs.~(\ref{eq:tophat_pk}) and~(\ref{eq:delp_uncorr})  that we need $W_\mathrm{2D}(|{\bf k}_\perp -{\bf q}_\perp| D_\mathrm{fc})$ for which we can use the addition theorem for 2D top-hat functions~\citep{BerColGaz02},
\beqa
W_\mathrm{2D}(|{\bf k}_\perp -{\bf q}_\perp| D_\mathrm{fc}) &=& \sum_{k=0} (k+1)\, U_k(\hat{k}_\perp\cdot \hat{q}_\perp) \nonumber
\\ & & W_\mathrm{2D}^{(k/2)}(k_\perp D_\mathrm{fc}) \,
W_\mathrm{2D}^{(k/2)}(q_\perp D_\mathrm{fc}) \nonumber \\ & & 
\label{ADDtheo}
\eeqa
where the $U_k$'s are the Chebyshev polynomials and $W_\mathrm{2D}^{(k/2)}(x) \equiv 2J_{k+1}(x)/x$. Now, again, as we are interested in scales for which $k D_\mathrm{fc} \ll 1$ is an excellent approximation, dropping ${\cal O}(k_\perp D_\mathrm{fc})^2$ we can just use the $k=0$ term in this expression. This gives us $W_\mathrm{2D}(|{\bf k}_\perp -{\bf q}_\perp| D_\mathrm{fc}) \approx W_\mathrm{2D}( q_\perp D_\mathrm{fc})$ as expected and leads to,
\beqa
\Delta P^\mathrm{corr}({\bf k}) &\approx& -{f_s \pi D_\mathrm{fc}^2}\int {d^2q_\perp\over (2\pi)^2} \, P(k_\parallel,q_\perp) \, W_\mathrm{2D}( q_\perp D_\mathrm{fc}) \nonumber \\ & &  \label{eq:delp_corr}
\eeqa
This is a simple result, showing that the correlated effect of fiber collisions is  proportional to the effective survey area affected by fiber collisions and to the  integral of the power spectrum over 2D modes perpendicular to the line of sight smoothed at the fiber collision scale. 
The multipole components of Eq.~(\ref{eq:delp_corr}) are, after expanding $P(k_\parallel,q_\perp)$ in multipoles,
\beq
\Delta P^\mathrm{corr}_l(k) \approx -\frac{f_s D_\mathrm{fc}^2}{2} \sum_{l'=0}^\infty \int_0^\infty q dq P_{l'}(q) \, f_{l l'}(k,q), 
\label{DeltaPell}
\eeq
where, again neglecting ${\cal O}(k D_\mathrm{fc})^2$,
\beqa
f_{ll'}(k,q) &\equiv &
\Big(\frac{2l+1}{2}\Big) \int_{\mathrm{max}(-1,-q/k)}^{\mathrm{min}(1,q/k)} d\mu  \, {\cal L}_l(\mu)\, {\cal L}_{l'}(k\mu/q) \nonumber \\ & &
 \  \ \ \ \ \ \ \ \ \ \ \ \ \times \ W_\mathrm{2D}( q\, D_\mathrm{fc})
\label{fellellp}
\eeqa
This has a simple expression for $l=l'$,
\beq
f_{ll}(k,q) = f_*(k,q)\, W_\mathrm{2D}( q\, D_\mathrm{fc})\, \Big(\frac{k_<}{k_>}\Big)^l
\label{fdiag}
\eeq
where $f_*(k,q)=q/k$ for $q\leq k$ and unity otherwise, and $k_>=\mathrm{max}(k,q)$ and $k_<=\mathrm{min}(k,q)$. On the other hand, off the diagonal we have ($l \neq l'$)
\beq
f_{ll'}(k,q) = f_*(k,q)\, W_\mathrm{2D}( q\, D_\mathrm{fc})\, \Big(\frac{2l+1}{2}\Big) \,H_{l_>l_<}\Big(\frac{k_<}{k_>}\Big),
\label{foffdiag}
\eeq
where $l_{>}=\mathrm{max}(l,l')$ and similarly $l_<$, and  $H_{l_>l_<}(x)$ is a polynomial of degree $l_>$ which vanishes unless $l$ and $k$ are both larger or  smaller than $l'$ and $q$ respectively. The first few polynomials are listed in the Appendix. Since $f_{l>l'}(k<q) = f_{l<l'}(k>q)=0$ it is convenient to split the integrals depending on whether $q$ is larger or smaller than $k$, which leads to
\beqa
\Delta P^\mathrm{corr}_l(k) &\approx & -\frac{f_s D_\mathrm{fc}^2}{2} \Bigg[\,
\sum_{l'\leq l} \int_0^k q dq \, P_{l'}(q) \, f_{l l'}(q\leq k) \nonumber \\ & & + 
\sum_{l'\geq l} \int_k^\infty q dq\, P_{l'}(q) \, f_{l l'}(q\geq k) \Bigg],
\label{DeltaPell2}
\eeqa
which shows that the change of power spectrum multipole $l$ due to correlated 
fiber collisions comes from long modes of lower multipoles ($l'\leq l$) and 
short modes of higher multipoles ($l'\geq l$). Going back to the results 
displayed in Figure~\ref{fig:2pcf_tophat}, we can now formulate how our results 
change if we use the observed numerical profile in the right panel of 
Figure~\ref{fig:2pcf_tophat} (red line) instead of the top-hat (black dashed). 
One can check that to leading order in $k D_\mathrm{fc}$, which is all we are using in this paper, our expression for the uncorrelated and correlated change in power are valid as long as we replace the 2D top-hat by the numerical profile in Eq.~(\ref{fellellp}), and redefine the scale $D_\mathrm{fc}$ that appears in Eqs.~(\ref{DeltaPuncorrexp}) and~(\ref{DeltaPell}) from the area of the numerical profile, that is 
\beq
\int d^2r_\perp \, W_\mathrm{2D}({\bf r}_\perp) \equiv \pi\, D_\mathrm{fc}^2
\label{redefDfc}
\eeq

\begin{figure*}
\begin{center}
\includegraphics[scale=0.5]{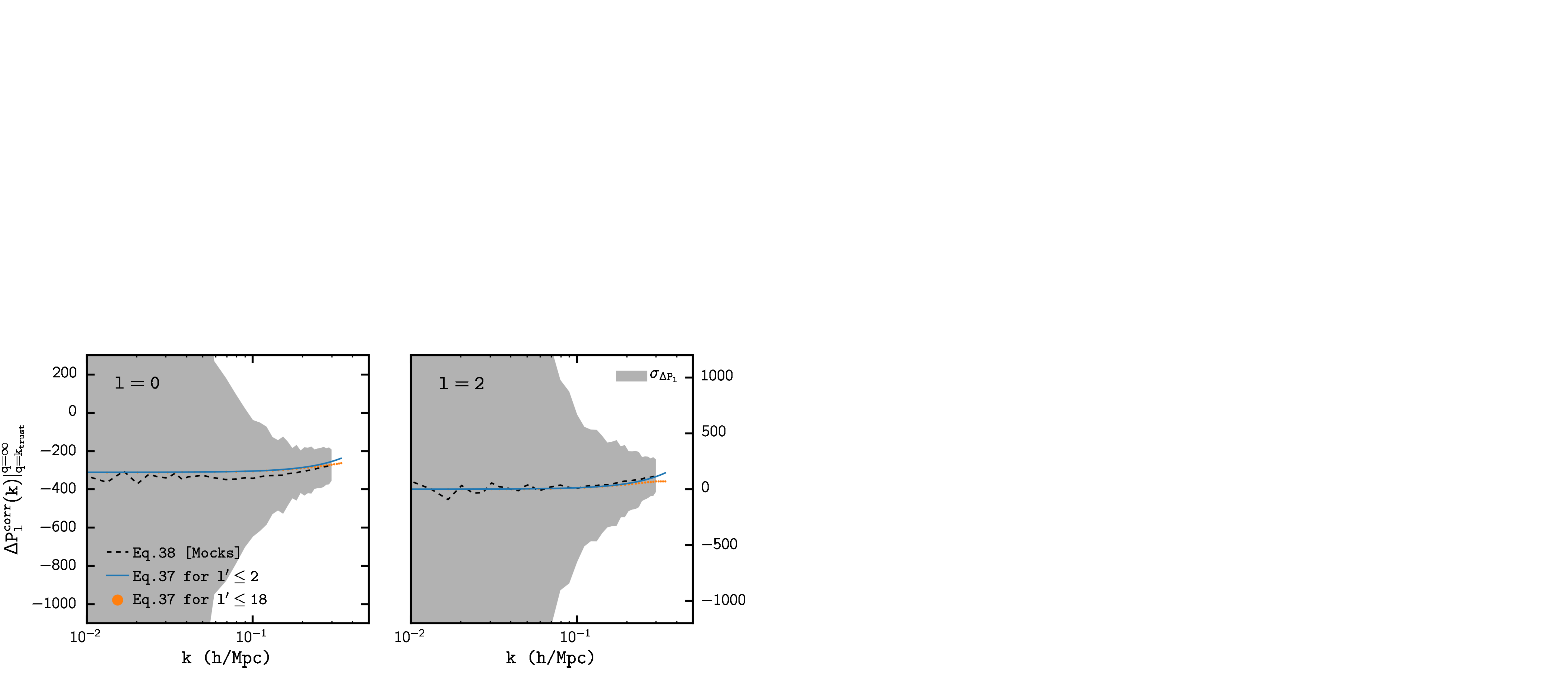}
\caption{Comparison of the correlated power spectrum residuals from unreliable modes obtained from mocks (dashed), Eq.~(\ref{eq:delp_untrust_nseries}), 
to the polynomial approximation of  
Eq.~(\ref{eq:delp_poly}) for $l' \leq 18$ (orange). The left and right panels
correspond to $l = 0$ and $2$ respectively. The gray shaded region is the 
standard deviation for the Nseries $(P_l^\mathrm{NN} - P_l^\mathrm{true})$.  
We also include Eq.~(\ref{eq:delp_poly}) evaluated only for $l' \leq 2$ (blue). 
The agreement between Eq.~(\ref{eq:delp_poly}) for $l' \leq 2$ and Eq.~(\ref{eq:delp_untrust_nseries}) demonstrate that while higher orders of $l'$ are 
necessary to properly model $\Delta P_l^\mathrm(k)$ at higher $k$ values, 
for $k < k_\mathrm{trust}$ ($0.3\;h/\mathrm{Mpc}$ above) $l' \leq 2 $ are 
sufficient.}
\label{fig:delP_untrust}
\end{center}
\end{figure*}

We now proceed to testing these results, for which we need the true power spectrum multipoles down to small scales to feed into Eq.~(\ref{DeltaPell2}). Unfortunately, in the nonlinear regime the multipole expansion is not very efficient (in the sense that the amplitude of multipoles does not decrease sharply with increasing multipole), so a large number of multipoles $l'$ is required to capture the contribution from small scale modes.  Measuring multipoles higher than the hexadecapole for realistic survey geometries  using our estimator becomes expensive due to the number of 
Fast Fourier Transforms (FFTs) that needs to be computed, and even for the most efficient version of the multipole estimators that requires only 7 FFTs one would worry about increased cosmic variance (see discussion in \citealt{Scoccimarro:2015aa}). 

A more efficient approach is to use the Nseries simulation boxes to test Eqs.~(\ref{eq:delp_uncorr}) and~(\ref{DeltaPell2}). The Nseries simulation boxes are the 
original simulations where the Nseries mocks were 
cut out from (Section \ref{sec:catalog}). Since the Nseries mocks 
are cut outs of the boxes, 
discrepancies in their power spectra are caused by the BOSS survey 
geometry and occur mainly at the largest scales, $k < 0.05\;h/\mathrm{Mpc}$ \citep{Beutler:2014aa,Grieb:2016aa}. 
At smaller scales, the difference between the power spectrum monopole, 
quadrupole and hexadecapole of Nseries mocks versus the Nseries boxes are 
negligible. 
Therefore, we calculate the $P_{l'}(q)$ from the Nseries simulation box, 
using periodic boundary conditions, which only requires one FFT and go up to $q = 43.5\;h/\mathrm{Mpc}$ and $l'=18$ to compute the corrections predicted by  Eq.~(\ref{DeltaPell2}).  

In Figure \ref{fig:delP}, we compare $\Delta P_l = \Delta P_l^\mathrm{corr} + 
\Delta P_l^\mathrm{uncorr}$ calculated from the Nseries Box power spectrum multipoles using Eqs.~(\ref{eq:delp_uncorr}) and~(\ref{DeltaPell2}) (\nseriescolor) to the Nseries mock catalogs power spectrum residuals, $\Delta P_l = P_l^\mathrm{NN} - P_l^\mathrm{true}$ (dashed). 
The left panel compares the monopoles ($l = 0$) while the right panel compares 
the quadrupoles ($l=2$). We also include in the gray shaded 
region, the standard deviation of Nseries mock catalogs power spectrum residuals,
$\sigma_{\Delta P_l}$. For both the monopole and quadrupole, the predictions (orange) agree with the measured residuals from NN-corrected fiber collisions (dashed black) well within the errors throughout the probed $k$ range up to $k=0.83\;h/\mathrm{Mpc}$. At low-$k$, the downturn (upturn) in the monopole (quadrupole) is due to the contribution of the $k^{-1}$ uncorrelated piece. 
The overall quality of the  agreement demonstrates that the effective window method can be used to robustly 
estimate the effect of fiber collisions on $P_l(k)$. Furthermore, with its
excellent performance for the quadrupole, the effective window approach 
provides an improvement over the LOS reconstruction method (Section
\ref{sec:dlospeak}).

\subsubsection{In Practice} \label{sec:tophat_practice}
There are, however, practical limitations to the effective window model 
as it described above. The 
$\Delta P^\mathrm{corr}_l$ calculations in 
Eq.~(\ref{DeltaPell2}) involves integrating the power spectrum over the $q$
range of $0$ to $\infty$. While this integral converges for $q \approx10\;h/\mathrm{Mpc}$ 
for both monopole and quadrupole, in practice one cannot compute reliably the power spectrum multipoles down to these scales. We now 
discuss a way to overcome this issue.

Let $k_\mathrm{trust}$ represent the scale up to which we can calculate reliably power 
spectrum multipoles. We therefore split the second term in Eq.~(\ref{DeltaPell2}) 
into a reliable piece (integration from $k$ to $k_\mathrm{trust}$) and an 
unreliable piece (integration from $k_\mathrm{trust}$ to $\infty$), so schematically
\beq \label{eq:delp_split}
\Delta P^\mathrm{corr}_l =  \Delta P^\mathrm{corr}_l \bigg|_{q=0}^{q=k_\mathrm{trust}} +  
\Delta P^\mathrm{corr}_l \bigg|_{q=k_\mathrm{trust}}^{q=\infty}.
\eeq
The first term can be reliably calculated from first principles 
since it involves modes from $q=0$ to $q=k_\mathrm{trust}$ and corresponds to  the first term  plus the reliable piece of the second term in Eq.~(\ref{DeltaPell2}). Now, the key fact is that because the second term in Eq.~(\ref{eq:delp_split}) only depends on $k$ through $f_{l l'}(q\geq k)$, from Eqs.~(\ref{fdiag}-\ref{foffdiag}) it follows that the $k$-dependence of the unreliable term is simply a polynomial in $k$,

\begin{align} \label{eq:delp_poly}
\Delta P^\mathrm{corr}_l \bigg|_{q=k_\mathrm{trust}}^{q=\infty} &= 
\sum\limits_{n= 0, 2, 4 ... } C_{l,n}\;k^{n}. 
\end{align}
The coefficients of the polynomial, $C_{l, n}$, are obtained by collecting powers of $k$ from the sum over the $H$-polynomial contributions to the second term in Eq.~(\ref{DeltaPell2}). How important are these unreliable contributions? In order to test this, in Figure \ref{fig:delP_untrust} 
we calculate $C_{l, n}$ from the $P_{l'}(q)$ multipoles measured 
from the Nseries simulation boxes (blue and orange for terms up 
to $l'=2$ and $18$ respectively) and compare to (black dashed)
\beq \label{eq:delp_untrust_nseries}
\Delta P_l^\mathrm{Nseries}(k) - \Delta P^\mathrm{uncorr}_l(k) - 
\Delta P^\mathrm{corr}_l(k) \bigg|^{q=k_\mathrm{trust}}_{q=0}
\eeq  
 where $\Delta P_l^\mathrm{Nseries}$ is the power spectrum 
residual $P_l^\mathrm{NN}- P_l^\mathrm{true}$ for the Nseries mocks (Figure \ref{fig:delP}). We once again include the standard deviation 
of the power spectrum residual in shaded gray. The agreement 
between Eq.~(\ref{eq:delp_poly}) and Eq.~(\ref{eq:delp_untrust_nseries}) 
is more or less equivalent to the agreement seen in Figure~\ref{fig:delP}, which includes uncorrelated and reliable correlated contributions as well; this 
should of course not come as a surprise.

More importantly, when we examine the contribution to 
$\Delta P^\mathrm{corr}_l |_{k_\mathrm{trust}}^\infty$ from each individual 
$l'$ order term of the Eq.~(\ref{eq:delp_poly}) polynomial, we find that the
main contributors at $k < k_\mathrm{trust} \sim 0.3\; h/\mathrm{Mpc}$ are 
the $l' \leq 2$ order terms. In fact, the higher order ($l' > 2$) terms of the 
polynomial contribute at higher $k$. For instance, the $l' = 4, 6,$ and $ 8$ terms 
only begin to significantly contribute at scales of $k > 0.3, \; 0.45$, and 
$0.6\;h/\mathrm{Mpc}$ respectively, which is not surprising since higher $k$ powers come together with increasing inverse powers of $q$ and thus suppress the value of the coefficients that result from integrating over small-scale modes. Hence, when we plot Eq.~(\ref{eq:delp_poly})  
for just $l' \leq 2$ (blue) in Figure~\ref{fig:delP_untrust}, we find that 
it is in good agreement with both Eq.~(\ref{eq:delp_poly}) for $l' \leq 18$ 
and Eq.~(\ref{eq:delp_untrust_nseries}). We also note that for $l = 2$, 
$C_{2, l'=0} = 0$ so the main contribution to 
$\Delta P^\mathrm{corr}_2(k < k_\mathrm{trust}) |_{k_\mathrm{trust}}^\infty$ 
comes solely from the $l' = 2$ term of the polynomial. 

To use the effective window method for cosmological inference, we can 
utilize the fact that Eq.~(\ref{eq:delp_poly}) with only $l' \leq 2$ terms 
provides an accurate estimate of the unreliable correlated change in power (Figure~\ref{fig:delP_untrust}). In cosmological analyses, the coefficients $C_{l, 0}$ and $C_{l, 2}$
can be nuisance parameters with priors obtained from  
mock catalogs. More specifically, for the quadrupole, since $C_{2, 0} = 0$
only one nuisance parameter is necessary. Meanwhile 
for the monopole, a constant shot noise term is typically already included as
a nuisance parameter in the analysis (\citealt{Beutler:2014aa,Beutler:2016aa,Grieb:2016aa,Gil-Marin:2016aa}) so there is 
also only one extra nuisance parameter for $l=0$. Therefore, 
by adding $C_{l, 2}$ as nuisance parameters to cosmological inference 
analyses of the power spectrum multipoles, we can use the effective window 
method to robustly marginalize over the effects of fiber collision for the 
entire $k$ range of power spectrum models based on perturbation theory. 

\section{Summary and Conclusions} \label{sec:summary}
Using simulated mock catalogs designed specifically for interpreting BOSS 
clustering measurements with realistically imposed fiber collisions, we 
demonstrate that the Nearest Neighbor method (NN), most common used for dealing 
with fiber collisions, is insufficient in accounting for the effect of 
fiber collisions on the galaxy power spectrum monopole and quadrupole.  
Although fiber collisions have little significant effect on 
the power spectrum at large scales, their effect quickly overtakes sample 
variance on scales smaller than $k \approx 0.1 \;h/\mathrm{Mpc}$. At  $k \sim 0.3 \;h/\mathrm{Mpc}$
 fiber collisions have over a $7.3\%$ and $73\%$ 
impact on the power spectrum monopole and quadrupole, respectively. The 
effect is equivalent to $7.3$ and $2.5$ times the sample variance of CMASS for 
$\delta k \approx 0.01\;h/\mathrm{Mpc}$, leading to a binning-independent scale of validity of the NN method of 
$k_{\chi^2}=0.068\;h/\mathrm{Mpc}$ for the monopole and $k_{\chi^2}=0.17\;h/\mathrm{Mpc}$ for the 
quadrupole (see bottom panel of Figure~\ref{fig:dlospeak_norm_resid}).  
Consequently at these scales, measurements of the power spectrum becomes 
dominated by the systematic effects of fiber collisions. 

Some recent methods (\citealt{Beutler:2014aa,Gil-Marin:2014aa,Beutler:2016aa,Grieb:2016aa,Gil-Marin:2016aa}) have supplemented
the NN method with adjustments to the constant shot noise term in the power spectrum 
estimator. While these methods improve the overall residual for the monopole, e.g. $k_{\chi^2}=0.17\;h/\mathrm{Mpc}$ for the method by \cite{Gil-Marin:2014aa}, they fail to account for the $k$-dependence of the systematic effect on smaller 
scales. Furthermore, since the quadrupole does not have a shot 
noise term, these methods provide no improvements for $l \geq 2$. 

In this paper, we first model the distribution of the line-of-sight displacement between 
fiber collided pairs using  mock catalogs. From the model, we statistically reconstruct the 
clustering of fiber collided galaxies that reside in the same halo. This, combined with the actual shot noise subtraction   
of the power spectrum estimator that accounts for chance alignments, 
leads to our LOS Reconstruction method that recovers very well the true power 
spectrum monopole from fiber collided data. As an added advantage, the method 
only relies on parameters ($\sigma_\mathrm{LOS}$ and $f_\mathrm{peak}$) 
measured from the actual observations. This makes the performance of the method 
independent from the accuracy of the mock catalogs, which are known to be unreliable 
at small scales. 

Using the LOS Reconstruction method, we can recover the true power 
spectrum monopole to scales well beyond previous methods. The LOS Reconstruction monopole power spectrum residuals remain within sample 
variance until $k \sim 0.53\;h/\mathrm{Mpc}$ and $k_{\chi^2}$ extends to  
$0.29\;h/\mathrm{Mpc}$. However, for the power spectrum quadrupole
at $k = 0.2\;h/\mathrm{Mpc}$,
the LOS Reconstruction method only reduces the discrepancy between the 
fiber collided $P_2(k)$ and the true $P_2(k)$ to roughly the sample variance. 
Therefore, the true monopole power spectrum estimate from 
the LOS reconstruction method can be compared to the systematics free predicted  
power spectrum monopole to infer the cosmological parameters of interest without 
biases from fiber collisions, but for the quadrupole power spectrum the method is not a substantial improvement over previous methods. We trace this problem  to the fact that the quadrupole is more sensitive to the object by object finger of god effect, while the LOS reconstruction works only statistically starting from the distribution of close pairs. 

To improve on the LOS reconstruction results we develop the effective window method which, rather than attempting to correct the data before making measurements, computes theoretical predictions of the fiber-collided power spectrum multipoles.
 In this approach, we approximate the effect that 
fiber collisions have on the two-dimensional configuration space two-point 
correlation function of the NN method as a scaled top-hat function. 
Then the effect of fiber collisions can be written as the sum of two contributions: 1) that of uncorrelated chance collisions, with an amplitude proportional to the 
the effective survey area affected by fiber collisions times the wavelength of perturbations, and 2) that of correlated collisions, which is also proportional to the effective survey area affected by fiber collisions and to the integral of the power spectrum over 2D modes perpendicular to the line of sight smoothed at the fiber collision scale.

Using high resolution mock catalogs, we demonstrate that our  analytic prescription 
 accurately models the power spectrum residuals from the NN method to within sample variance of BOSS volumes  
at $k < 0.83\;h/\mathrm{Mpc}$ for both the monopole and quadrupole when the true power spectrum is known down to small scales from simulations, allowing to compute the fiber-collided predictions. Since typically we do not have fast reliable ways of computing the small scale power spectrum, we develop a practical approach when the power spectrum predictions are reliable up to some scale $k_\mathrm{trust}$. We split the contributions of the correlated fiber collisions effect into a  piece that can be calculated reliably as it depends on large-scale modes, and an unreliable piece that depends on modes that are not under control. We show that the latter piece can be written as polynomials in $k$, and demonstrate that for scales up to $k \sim 0.3\;h/\mathrm{Mpc}$, the unreliable contribution can be accurately estimated by a quadratic  polynomial in $k$. In principle, this method can be applied to larger $k_\mathrm{trust}$ than used here as a reasonable example ($k_\mathrm{trust} = 0.3\;h/\mathrm{Mpc}$). 

Therefore, using the effective window method we can model the fiber collided power 
spectrum as the systematics-free  power spectrum plus three contributions due to  
fiber collisions: an uncorrelated piece (independent of the model power spectrum), a calculable piece (which involves integrating the model power spectrum over 2D long-wavelength modes perpendicular to the line of sight), and an unreliable contribution that is a quadratic polynomial, $C_{l,0} + C_{l,2}\, k^2$. 
While the precise values of $C_{l, n}$ cannot be robustly predicted in practice 
because of its dependence on small scale power, the coefficients 
can be treated as nuisance parameters in the analysis. Typically a constant shot 
noise term is already included as a nuisance parameter, while the constant contribution vanishes for higher multipoles, therefore only one extra parameter 
per multipole is required (the $k^2$ corrections). For cosmological parameter inference, the fiber collided model power spectrum 
can be compared directly to the observed fiber 
collided power spectrum. Then by marginalizing over these free coefficients, we marginalize 
over the effect of small-scale power induced fiber collisions on the power spectrum, which allows us to robustly 
infer the cosmological parameters of interest.

The fiber collision correction methods we present will enable us to robustly 
account for the effects of fiber collisions in galaxy clustering analyses 
to the smallest scales allowed by theoretical predictions. They can also be extended to 
future surveys such as eBOSS  or any other large fiber-fed 
surveys that suffer from systematic effects of fiber collisions. Our fiber
collision correction method can also be extended to higher order clustering 
statistics such as bispectrum (Hahn et al., in prep.). We will use the methods 
presented in this paper to analyze the galaxy power spectrum and bispectrum 
multipoles in future work.

\bigskip

\section*{Acknowledgements}
CHH and MRB were supported by NSF-AST-1109432 and NSF-AST-1211644.
SRT is grateful for support from the Campus de Excelencia Internacional UAM/CSIC.
We thank A. I. Malz, Mohammadjavad Vakili, Johan Comparat and particularly 
David W. Hogg for helpful discussions. CHH also thanks the Instituto 
de F\'{i}sica Teo\'{o}rica (UAM/CSIC) and 
particularly Francisco Prada for their hospitality during his summer 
visit, where part of this work was completed.

\bibliographystyle{yahapj}
\bibliography{fc_paper}

\begin{thebibliography}{}
\providecommand\natexlab[1]{#1}
\providecommand\JournalTitle[1]{#1}

\bibitem[{{Alonso}(2012)}]{Alonso:2012aa}
{Alonso}, D. 2012, \JournalTitle{ArXiv e-prints},
  \href{http://arxiv.org/abs/1210.1833}{{\sffamily arXiv:1210.1833
  [astro-ph.IM]}}

\bibitem[{{Anderson} {et~al.}(2012){Anderson}, {Aubourg}, {Bailey}, {Bizyaev},
  {Blanton}, {Bolton}, {Brinkmann}, {Brownstein}, {Burden}, {Cuesta}, {da
  Costa}, {Dawson}, {de Putter}, {Eisenstein}, {Gunn}, {Guo}, {Hamilton},
  {Harding}, {Ho}, {Honscheid}, {Kazin}, {Kirkby}, {Kneib}, {Labatie},
  {Loomis}, {Lupton}, {Malanushenko}, {Malanushenko}, {Mandelbaum}, {Manera},
  {Maraston}, {McBride}, {Mehta}, {Mena}, {Montesano}, {Muna}, {Nichol},
  {Nuza}, {Olmstead}, {Oravetz}, {Padmanabhan}, {Palanque-Delabrouille}, {Pan},
  {Parejko}, {P{\^a}ris}, {Percival}, {Petitjean}, {Prada}, {Reid}, {Roe},
  {Ross}, {Ross}, {Samushia}, {S{\'a}nchez}, {Schlegel}, {Schneider},
  {Sc{\'o}ccola}, {Seo}, {Sheldon}, {Simmons}, {Skibba}, {Strauss}, {Swanson},
  {Thomas}, {Tinker}, {Tojeiro}, {Maga{\~n}a}, {Verde}, {Wagner}, {Wake},
  {Weaver}, {Weinberg}, {White}, {Xu}, {Y{\`e}che}, {Zehavi}, \&
  {Zhao}}]{Anderson:2012aa}
{Anderson}, L., {Aubourg}, E., {Bailey}, S., {et~al.} 2012,
  \href{http://dx.doi.org/10.1111/j.1365-2966.2012.22066.x}{\JournalTitle{\mnras},
  427, 3435}

\bibitem[{{Behroozi} {et~al.}(2013){Behroozi}, {Wechsler}, \&
  {Wu}}]{Behroozi:2013aa}
{Behroozi}, P.~S., {Wechsler}, R.~H., \& {Wu}, H.-Y. 2013,
  \href{http://dx.doi.org/10.1088/0004-637X/762/2/109}{\JournalTitle{\apj},
  762, 109}

\bibitem[{{Berlind} {et~al.}(2006){Berlind}, {Frieman}, {Weinberg}, {Blanton},
  {Warren}, {Abazajian}, {Scranton}, {Hogg}, {Scoccimarro}, {Bahcall},
  {Brinkmann}, {Gott}, {Kleinman}, {Krzesinski}, {Lee}, {Miller}, {Nitta},
  {Schneider}, {Tucker}, {Zehavi}, \& {SDSS Collaboration}}]{Berlind:2006aa}
{Berlind}, A.~A., {Frieman}, J., {Weinberg}, D.~H., {et~al.} 2006,
  \href{http://dx.doi.org/10.1086/508170}{\JournalTitle{\apjs}, 167, 1}

\bibitem[{{Beutler} {et~al.}(2014){Beutler}, {Saito}, {Seo}, {Brinkmann},
  {Dawson}, {Eisenstein}, {Font-Ribera}, {Ho}, {McBride}, {Montesano},
  {Percival}, {Ross}, {Ross}, {Samushia}, {Schlegel}, {S{\'a}nchez}, {Tinker},
  \& {Weaver}}]{Beutler:2014aa}
{Beutler}, F., {Saito}, S., {Seo}, H.-J., {et~al.} 2014,
  \href{http://dx.doi.org/10.1093/mnras/stu1051}{\JournalTitle{\mnras}, 443,
  1065}

\bibitem[{{Beutler} {et~al.}(2016){Beutler}, {Seo}, {Saito}, {Chuang},
  {Cuesta}, {Eisenstein}, {Gil-Mar{\'{\i}}n}, {Grieb}, {Hand}, {Kitaura},
  {Modi}, {Nichol}, {Olmstead}, {Percival}, {Prada}, {S{\'a}nchez},
  {Rodriguez-Torres}, {Ross}, {Ross}, {Schneider}, {Tinker}, {Tojeiro}, \&
  {Vargas-Maga{\~n}a}}]{Beutler:2016aa}
{Beutler}, F., {Seo}, H.-J., {Saito}, S., {et~al.} 2016, \JournalTitle{ArXiv
  e-prints}, \href{http://arxiv.org/abs/1607.03150}{{\sffamily
  arXiv:1607.03150}}

\bibitem[{{Bianchi} {et~al.}(2016){Bianchi}, {Percival}, \&
  {Bel}}]{Bianchi:2016aa}
{Bianchi}, D., {Percival}, W., \& {Bel}, J. 2016, \JournalTitle{ArXiv
  e-prints}, \href{http://arxiv.org/abs/1602.02780}{{\sffamily
  arXiv:1602.02780}}

\bibitem[{{Carretero} {et~al.}(2015){Carretero}, {Castander}, {Gazta{\~n}aga},
  {Crocce}, \& {Fosalba}}]{Carretero:2015aa}
{Carretero}, J., {Castander}, F.~J., {Gazta{\~n}aga}, E., {Crocce}, M., \&
  {Fosalba}, P. 2015,
  \href{http://dx.doi.org/10.1093/mnras/stu2402}{\JournalTitle{\mnras}, 447,
  646}

\bibitem[{{Chuang} {et~al.}(2015){Chuang}, {Zhao}, {Prada}, {Munari}, {Avila},
  {Izard}, {Kitaura}, {Manera}, {Monaco}, {Murray}, {Knebe}, {Sc{\'o}ccola},
  {Yepes}, {Garcia-Bellido}, {Mar{\'{\i}}n}, {M{\"u}ller}, {Skibba}, {Crocce},
  {Fosalba}, {Gottl{\"o}ber}, {Klypin}, {Power}, {Tao}, \&
  {Turchaninov}}]{Chuang:2015aa}
{Chuang}, C.-H., {Zhao}, C., {Prada}, F., {et~al.} 2015,
  \href{http://dx.doi.org/10.1093/mnras/stv1289}{\JournalTitle{\mnras}, 452,
  686}

\bibitem[{{Cole} {et~al.}(1998){Cole}, {Hatton}, {Weinberg}, \&
  {Frenk}}]{Cole:1998aa}
{Cole}, S., {Hatton}, S., {Weinberg}, D.~H., \& {Frenk}, C.~S. 1998,
  \href{http://dx.doi.org/10.1046/j.1365-8711.1998.01936.x}{\JournalTitle{\mnras},
  300, 945}

\bibitem[{{Colless}(1999)}]{Colless:1999aa}
{Colless}, M. 1999,
  \href{http://dx.doi.org/10.1098/rsta.1999.0317}{\JournalTitle{Royal Society
  of London Philosophical Transactions Series A}, 357, 105}

\bibitem[{{Cuesta} {et~al.}(2016){Cuesta}, {Vargas-Maga{\~n}a}, {Beutler},
  {Bolton}, {Brownstein}, {Eisenstein}, {Gil-Mar{\'{\i}}n}, {Ho}, {McBride},
  {Maraston}, {Padmanabhan}, {Percival}, {Reid}, {Ross}, {Ross}, {S{\'a}nchez},
  {Schlegel}, {Schneider}, {Thomas}, {Tinker}, {Tojeiro}, {Verde}, \&
  {White}}]{Cuesta:2016aa}
{Cuesta}, A.~J., {Vargas-Maga{\~n}a}, M., {Beutler}, F., {et~al.} 2016,
  \href{http://dx.doi.org/10.1093/mnras/stw066}{\JournalTitle{\mnras}, 457,
  1770}

\bibitem[{{Dawson} {et~al.}(2013){Dawson}, {Schlegel}, {Ahn}, {Anderson},
  {Aubourg}, {Bailey}, {Barkhouser}, {Bautista}, {Beifiori}, {Berlind},
  {Bhardwaj}, {Bizyaev}, {Blake}, {Blanton}, {Blomqvist}, {Bolton}, {Borde},
  {Bovy}, {Brandt}, {Brewington}, {Brinkmann}, {Brown}, {Brownstein}, {Bundy},
  {Busca}, {Carithers}, {Carnero}, {Carr}, {Chen}, {Comparat}, {Connolly},
  {Cope}, {Croft}, {Cuesta}, {da Costa}, {Davenport}, {Delubac}, {de Putter},
  {Dhital}, {Ealet}, {Ebelke}, {Eisenstein}, {Escoffier}, {Fan}, {Filiz Ak},
  {Finley}, {Font-Ribera}, {G{\'e}nova-Santos}, {Gunn}, {Guo}, {Haggard},
  {Hall}, {Hamilton}, {Harris}, {Harris}, {Ho}, {Hogg}, {Holder}, {Honscheid},
  {Huehnerhoff}, {Jordan}, {Jordan}, {Kauffmann}, {Kazin}, {Kirkby}, {Klaene},
  {Kneib}, {Le Goff}, {Lee}, {Long}, {Loomis}, {Lundgren}, {Lupton}, {Maia},
  {Makler}, {Malanushenko}, {Malanushenko}, {Mandelbaum}, {Manera}, {Maraston},
  {Margala}, {Masters}, {McBride}, {McDonald}, {McGreer}, {McMahon}, {Mena},
  {Miralda-Escud{\'e}}, {Montero-Dorta}, {Montesano}, {Muna}, {Myers},
  {Naugle}, {Nichol}, {Noterdaeme}, {Nuza}, {Olmstead}, {Oravetz}, {Oravetz},
  {Owen}, {Padmanabhan}, {Palanque-Delabrouille}, {Pan}, {Parejko},
  {P{\^a}ris}, {Percival}, {P{\'e}rez-Fournon}, {P{\'e}rez-R{\`a}fols},
  {Petitjean}, {Pfaffenberger}, {Pforr}, {Pieri}, {Prada}, {Price-Whelan},
  {Raddick}, {Rebolo}, {Rich}, {Richards}, {Rockosi}, {Roe}, {Ross}, {Ross},
  {Rossi}, {Rubi{\~n}o-Martin}, {Samushia}, {S{\'a}nchez}, {Sayres}, {Schmidt},
  {Schneider}, {Sc{\'o}ccola}, {Seo}, {Shelden}, {Sheldon}, {Shen}, {Shu},
  {Slosar}, {Smee}, {Snedden}, {Stauffer}, {Steele}, {Strauss}, {Streblyanska},
  {Suzuki}, {Swanson}, {Tal}, {Tanaka}, {Thomas}, {Tinker}, {Tojeiro},
  {Tremonti}, {Vargas Maga{\~n}a}, {Verde}, {Viel}, {Wake}, {Watson}, {Weaver},
  {Weinberg}, {Weiner}, {West}, {White}, {Wood-Vasey}, {Yeche}, {Zehavi},
  {Zhao}, \& {Zheng}}]{Dawson:2013aa}
{Dawson}, K.~S., {Schlegel}, D.~J., {Ahn}, C.~P., {et~al.} 2013,
  \href{http://dx.doi.org/10.1088/0004-6256/145/1/10}{\JournalTitle{\aj}, 145,
  10}

\bibitem[{{Dawson} {et~al.}(2015){Dawson}, {Kneib}, {Percival}, {Alam},
  {Albareti}, {Anderson}, {Armengaud}, {Aubourg}, {Bailey}, {Bautista},
  {Berlind}, {Bershady}, {Beutler}, {Bizyaev}, {Blanton}, {Blomqvist},
  {Bolton}, {Bovy}, {Brandt}, {Brinkmann}, {Brownstein}, {Burtin}, {Busca},
  {Cai}, {Chuang}, {Clerc}, {Comparat}, {Cope}, {Croft}, {Cruz-Gonzalez}, {da
  Costa}, {Cousinou}, {Darling}, {de la Torre}, {Delubac}, {du Mas des
  Bourboux}, {Dwelly}, {Ealet}, {Eisenstein}, {Eracleous}, {Escoffier}, {Fan},
  {Finoguenov}, {Font-Ribera}, {Frinchaboy}, {Gaulme}, {Georgakakis}, {Green},
  {Guo}, {Guy}, {Ho}, {Holder}, {Huehnerhoff}, {Hutchinson}, {Jing}, {Jullo},
  {Kamble}, {Kinemuchi}, {Kirkby}, {Kitaura}, {Klaene}, {Laher}, {Lang},
  {Laurent}, {Le Goff}, {Li}, {Liang}, {Lima}, {Lin}, {Lin}, {Lin}, {Long},
  {Lundgren}, {MacDonald}, {Geimba Maia}, {Malanushenko}, {Malanushenko},
  {Mariappan}, {McBride}, {McGreer}, {Menard}, {Merloni}, {Meza},
  {Montero-Dorta}, {Muna}, {Myers}, {Nandra}, {Naugle}, {Newman}, {Noterdaeme},
  {Nugent}, {Ogando}, {Olmstead}, {Oravetz}, {Oravetz}, {Padmanabhan},
  {Palanque-Delabrouille}, {Pan}, {Parejko}, {Paris}, {Peacock}, {Petitjean},
  {Pieri}, {Pisani}, {Prada}, {Prakash}, {Raichoor}, {Reid}, {Rich}, {Ridl},
  {Rodriguez-Torres}, {Carnero Rosell}, {Ross}, {Rossi}, {Ruan}, {Salvato},
  {Sayres}, {Schneider}, {Schlegel}, {Seljak}, {Seo}, {Sesar}, {Shandera},
  {Shu}, {Slosar}, {Sobreira}, {Strauss}, {Streblyanska}, {Suzuki}, {Tao},
  {Tinker}, {Tojeiro}, {Vargas-Magana}, {Wang}, {Weaver}, {Weinberg}, {White},
  {Wood-Vasey}, {Yeche}, {Zhai}, {Zhao}, {Zhao}, {Zheng}, {Ben Zhu}, \&
  {Zou}}]{Dawson:2015aa}
{Dawson}, K.~S., {Kneib}, J.-P., {Percival}, W.~J., {et~al.} 2015,
  \JournalTitle{ArXiv e-prints},
  \href{http://arxiv.org/abs/1508.04473}{{\sffamily arXiv:1508.04473}}

\bibitem[{{Feldman} {et~al.}(1994){Feldman}, {Kaiser}, \&
  {Peacock}}]{Feldman:1994aa}
{Feldman}, H.~A., {Kaiser}, N., \& {Peacock}, J.~A. 1994,
  \href{http://dx.doi.org/10.1086/174036}{\JournalTitle{\apj}, 426, 23}

\bibitem[{{Gil-Mar{\'{\i}}n} {et~al.}(2014){Gil-Mar{\'{\i}}n}, {Nore{\~n}a},
  {Verde}, {Percival}, {Wagner}, {Manera}, \& {Schneider}}]{Gil-Marin:2014aa}
{Gil-Mar{\'{\i}}n}, H., {Nore{\~n}a}, J., {Verde}, L., {et~al.} 2014,
  \JournalTitle{ArXiv e-prints},
  \href{http://arxiv.org/abs/1407.5668}{{\sffamily arXiv:1407.5668}}

\bibitem[{{Gil-Mar{\'{\i}}n} {et~al.}(2016{\natexlab{a}}){Gil-Mar{\'{\i}}n},
  {Percival}, {Verde}, {Brownstein}, {Chuang}, {Kitaura},
  {Rodr{\'{\i}}guez-Torres}, \& {Olmstead}}]{Gil-Marin:2016ab}
{Gil-Mar{\'{\i}}n}, H., {Percival}, W.~J., {Verde}, L., {et~al.}
  2016{\natexlab{a}}, \JournalTitle{ArXiv e-prints},
  \href{http://arxiv.org/abs/1606.00439}{{\sffamily arXiv:1606.00439}}

\bibitem[{{Gil-Mar{\'{\i}}n} {et~al.}(2015){Gil-Mar{\'{\i}}n}, {Verde},
  {Nore{\~n}a}, {Cuesta}, {Samushia}, {Percival}, {Wagner}, {Manera}, \&
  {Schneider}}]{Gil-Marin:2015aa}
{Gil-Mar{\'{\i}}n}, H., {Verde}, L., {Nore{\~n}a}, J., {et~al.} 2015,
  \href{http://dx.doi.org/10.1093/mnras/stv1359}{\JournalTitle{\mnras}, 452,
  1914}

\bibitem[{{Gil-Mar{\'{\i}}n} {et~al.}(2016{\natexlab{b}}){Gil-Mar{\'{\i}}n},
  {Percival}, {Brownstein}, {Chuang}, {Grieb}, {Ho}, {Kitaura}, {Maraston},
  {Prada}, {Rodr{\'{\i}}guez-Torres}, {Ross}, {Samushia}, {Schlegel}, {Thomas},
  {Tinker}, \& {Zhao}}]{Gil-Marin:2016aa}
{Gil-Mar{\'{\i}}n}, H., {Percival}, W.~J., {Brownstein}, J.~R., {et~al.}
  2016{\natexlab{b}},
  \href{http://dx.doi.org/10.1093/mnras/stw1096}{\JournalTitle{\mnras}, 460,
  4188}

\bibitem[{{Grieb} {et~al.}(2016){Grieb}, {S{\'a}nchez}, {Salazar-Albornoz},
  {Scoccimarro}, {Crocce}, {Dalla Vecchia}, {Montesano}, {Gil-Mar{\'{\i}}n},
  {Ross}, {Beutler}, {Rodr{\'{\i}}guez-Torres}, {Chuang}, {Prada}, {Kitaura},
  {Cuesta}, {Eisenstein}, {Percival}, {Vargas-Magana}, {Tinker}, {Tojeiro},
  {Brownstein}, {Maraston}, {Nichol}, {Olmstead}, {Samushia}, {Seo},
  {Streblyanska}, \& {Zhao}}]{Grieb:2016aa}
{Grieb}, J.~N., {S{\'a}nchez}, A.~G., {Salazar-Albornoz}, S., {et~al.} 2016,
  \JournalTitle{ArXiv e-prints},
  \href{http://arxiv.org/abs/1607.03143}{{\sffamily arXiv:1607.03143}}

\bibitem[{{Guo} {et~al.}(2012){Guo}, {Zehavi}, \& {Zheng}}]{Guo:2012aa}
{Guo}, H., {Zehavi}, I., \& {Zheng}, Z. 2012,
  \href{http://dx.doi.org/10.1088/0004-637X/756/2/127}{\JournalTitle{\apj},
  756, 127}

\bibitem[{{Hamilton}(1997)}]{Hamilton:1997aa}
{Hamilton}, A.~J.~S. 1997,
  \href{http://dx.doi.org/10.1093/mnras/289.2.285}{\JournalTitle{\mnras}, 289,
  285}

\bibitem[{{Hogg}(1999)}]{Hogg:1999aa}
{Hogg}, D.~W. 1999, \JournalTitle{ArXiv Astrophysics e-prints},
  \href{http://arxiv.org/abs/astro-ph/9905116}{{\sffamily astro-ph/9905116}}

\bibitem[{{Howlett} {et~al.}(2015){Howlett}, {Manera}, \&
  {Percival}}]{Howlett:2015aa}
{Howlett}, C., {Manera}, M., \& {Percival}, W.~J. 2015,
  \href{http://dx.doi.org/10.1016/j.ascom.2015.07.003}{\JournalTitle{Astronomy
  and Computing}, 12, 109}

\bibitem[{{Izard} {et~al.}(2016){Izard}, {Crocce}, \& {Fosalba}}]{Izard:2016aa}
{Izard}, A., {Crocce}, M., \& {Fosalba}, P. 2016,
  \href{http://dx.doi.org/10.1093/mnras/stw797}{\JournalTitle{\mnras}, 459,
  2327}

\bibitem[{{Kitaura} {et~al.}(2016){Kitaura}, {Rodr{\'{\i}}guez-Torres},
  {Chuang}, {Zhao}, {Prada}, {Gil-Mar{\'{\i}}n}, {Guo}, {Yepes}, {Klypin},
  {Sc{\'o}ccola}, {Tinker}, {McBride}, {Reid}, {S{\'a}nchez},
  {Salazar-Albornoz}, {Grieb}, {Vargas-Magana}, {Cuesta}, {Neyrinck},
  {Beutler}, {Comparat}, {Percival}, \& {Ross}}]{Kitaura:2016aa}
{Kitaura}, F.-S., {Rodr{\'{\i}}guez-Torres}, S., {Chuang}, C.-H., {et~al.}
  2016, \href{http://dx.doi.org/10.1093/mnras/stv2826}{\JournalTitle{\mnras},
  456, 4156}

\bibitem[{{Klypin} {et~al.}(2014){Klypin}, {Yepes}, {Gottlober}, {Prada}, \&
  {Hess}}]{Klypin:2014aa}
{Klypin}, A., {Yepes}, G., {Gottlober}, S., {Prada}, F., \& {Hess}, S. 2014,
  \JournalTitle{ArXiv e-prints},
  \href{http://arxiv.org/abs/1411.4001}{{\sffamily arXiv:1411.4001}}

\bibitem[{{Landy} \& {Szalay}(1993)}]{Landy:1993aa}
{Landy}, S.~D., \& {Szalay}, A.~S. 1993,
  \href{http://dx.doi.org/10.1086/172900}{\JournalTitle{\apj}, 412, 64}

\bibitem[{{Makarem} {et~al.}(2014){Makarem}, {Kneib}, {Gillet}, {Bleuler},
  {Bouri}, {Jenni}, {Prada}, \& {Sanchez}}]{Makarem:2014aa}
{Makarem}, L., {Kneib}, J.-P., {Gillet}, D., {et~al.} 2014,
  \href{http://dx.doi.org/10.1051/0004-6361/201323202}{\JournalTitle{\aap},
  566, A84}

\bibitem[{{Manera} {et~al.}(2013){Manera}, {Scoccimarro}, {Percival},
  {Samushia}, {McBride}, {Ross}, {Sheth}, {White}, {Reid}, {S{\'a}nchez}, {de
  Putter}, {Xu}, {Berlind}, {Brinkmann}, {Maraston}, {Nichol}, {Montesano},
  {Padmanabhan}, {Skibba}, {Tojeiro}, \& {Weaver}}]{Manera:2013aa}
{Manera}, M., {Scoccimarro}, R., {Percival}, W.~J., {et~al.} 2013,
  \href{http://dx.doi.org/10.1093/mnras/sts084}{\JournalTitle{\mnras}, 428,
  1036}

\bibitem[{{Manera} {et~al.}(2015){Manera}, {Samushia}, {Tojeiro}, {Howlett},
  {Ross}, {Percival}, {Gil-Mar{\'{\i}}n}, {Brownstein}, {Burden}, \&
  {Montesano}}]{Manera:2015aa}
{Manera}, M., {Samushia}, L., {Tojeiro}, R., {et~al.} 2015,
  \href{http://dx.doi.org/10.1093/mnras/stu2465}{\JournalTitle{\mnras}, 447,
  437}

\bibitem[{{Maraston} {et~al.}(2013){Maraston}, {Pforr}, {Henriques}, {Thomas},
  {Wake}, {Brownstein}, {Capozzi}, {Tinker}, {Bundy}, {Skibba}, {Beifiori},
  {Nichol}, {Edmondson}, {Schneider}, {Chen}, {Masters}, {Steele}, {Bolton},
  {York}, {Weaver}, {Higgs}, {Bizyaev}, {Brewington}, {Malanushenko},
  {Malanushenko}, {Snedden}, {Oravetz}, {Pan}, {Shelden}, \&
  {Simmons}}]{Maraston:2013aa}
{Maraston}, C., {Pforr}, J., {Henriques}, B.~M., {et~al.} 2013,
  \href{http://dx.doi.org/10.1093/mnras/stt1424}{\JournalTitle{\mnras}, 435,
  2764}

\bibitem[{{Markwardt}(2009)}]{Markwardt:2009aa}
{Markwardt}, C.~B. 2009, in Astronomical Society of the Pacific Conference
  Series, Vol. 411, Astronomical Data Analysis Software and Systems XVIII, ed.
  D.~A. {Bohlender}, D.~{Durand}, \& P.~{Dowler}, 251

\bibitem[{{Monaco} {et~al.}(2013){Monaco}, {Sefusatti}, {Borgani}, {Crocce},
  {Fosalba}, {Sheth}, \& {Theuns}}]{Monaco:2013aa}
{Monaco}, P., {Sefusatti}, E., {Borgani}, S., {et~al.} 2013,
  \href{http://dx.doi.org/10.1093/mnras/stt907}{\JournalTitle{\mnras}, 433,
  2389}

\bibitem[{{Morales} {et~al.}(2012){Morales}, {Montero-Dorta}, {Azzaro},
  {Prada}, {S{\'a}nchez}, \& {Becerril}}]{Morales:2012aa}
{Morales}, I., {Montero-Dorta}, A.~D., {Azzaro}, M., {et~al.} 2012,
  \href{http://dx.doi.org/10.1111/j.1365-2966.2011.19774.x}{\JournalTitle{\mnras},
  419, 1187}

\bibitem[{{Munari} {et~al.}(2016){Munari}, {Monaco}, {Sefusatti}, {Castorina},
  {Mohammad}, {Anselmi}, \& {Borgani}}]{Munari:2016aa}
{Munari}, E., {Monaco}, P., {Sefusatti}, E., {et~al.} 2016, \JournalTitle{ArXiv
  e-prints}, \href{http://arxiv.org/abs/1605.04788}{{\sffamily
  arXiv:1605.04788}}

\bibitem[{{Okumura} {et~al.}(2015){Okumura}, {Hand}, {Seljak}, {Vlah}, \&
  {Desjacques}}]{Okumura:2015aa}
{Okumura}, T., {Hand}, N., {Seljak}, U., {Vlah}, Z., \& {Desjacques}, V. 2015,
  \href{http://dx.doi.org/10.1103/PhysRevD.92.103516}{\JournalTitle{\prd}, 92,
  103516}

\bibitem[{{Okumura} {et~al.}(2012){Okumura}, {Seljak}, \&
  {Desjacques}}]{Okumura:2012aa}
{Okumura}, T., {Seljak}, U., \& {Desjacques}, V. 2012,
  \href{http://dx.doi.org/10.1088/1475-7516/2012/11/014}{\JournalTitle{\jcap},
  11, 014}

\bibitem[{{Peebles}(1980)}]{Peebles:1980aa}
{Peebles}, P.~J.~E. 1980, {The large-scale structure of the universe}

\bibitem[{{Reid} {et~al.}(2012){Reid}, {Samushia}, {White}, {Percival},
  {Manera}, {Padmanabhan}, {Ross}, {S{\'a}nchez}, {Bailey}, {Bizyaev},
  {Bolton}, {Brewington}, {Brinkmann}, {Brownstein}, {Cuesta}, {Eisenstein},
  {Gunn}, {Honscheid}, {Malanushenko}, {Malanushenko}, {Maraston}, {McBride},
  {Muna}, {Nichol}, {Oravetz}, {Pan}, {de Putter}, {Roe}, {Ross}, {Schlegel},
  {Schneider}, {Seo}, {Shelden}, {Sheldon}, {Simmons}, {Skibba}, {Snedden},
  {Swanson}, {Thomas}, {Tinker}, {Tojeiro}, {Verde}, {Wake}, {Weaver},
  {Weinberg}, {Zehavi}, \& {Zhao}}]{Reid:2012aa}
{Reid}, B.~A., {Samushia}, L., {White}, M., {et~al.} 2012,
  \href{http://dx.doi.org/10.1111/j.1365-2966.2012.21779.x}{\JournalTitle{\mnras},
  426, 2719}

\bibitem[{{Rodr{\'{\i}}guez-Torres} {et~al.}(2015){Rodr{\'{\i}}guez-Torres},
  {Prada}, {Chuang}, {Guo}, {Klypin}, {Behroozi}, {Hahn}, {Comparat}, {Yepes},
  {Montero-Dorta}, {Brownstein}, {Maraston}, {McBride}, {Tinker},
  {Gottl{\"o}ber}, {Favole}, {Shu}, {Kitaura}, {Bolton}, {Scoccimarro},
  {Samushia}, {Schlegel}, {Schneider}, \& {Thomas}}]{Rodriguez-Torres:2015aa}
{Rodr{\'{\i}}guez-Torres}, S.~A., {Prada}, F., {Chuang}, C.-H., {et~al.} 2015,
  \JournalTitle{ArXiv e-prints},
  \href{http://arxiv.org/abs/1509.06404}{{\sffamily arXiv:1509.06404}}

\bibitem[{{Ross} {et~al.}(2012){Ross}, {Percival}, {S{\'a}nchez}, {Samushia},
  {Ho}, {Kazin}, {Manera}, {Reid}, {White}, {Tojeiro}, {McBride}, {Xu}, {Wake},
  {Strauss}, {Montesano}, {Swanson}, {Bailey}, {Bolton}, {Dorta}, {Eisenstein},
  {Guo}, {Hamilton}, {Nichol}, {Padmanabhan}, {Prada}, {Schlegel},
  {Maga{\~n}a}, {Zehavi}, {Blanton}, {Bizyaev}, {Brewington}, {Cuesta},
  {Malanushenko}, {Malanushenko}, {Oravetz}, {Parejko}, {Pan}, {Schneider},
  {Shelden}, {Simmons}, {Snedden}, \& {Zhao}}]{Ross:2012aa}
{Ross}, A.~J., {Percival}, W.~J., {S{\'a}nchez}, A.~G., {et~al.} 2012,
  \href{http://dx.doi.org/10.1111/j.1365-2966.2012.21235.x}{\JournalTitle{\mnras},
  424, 564}

\bibitem[{{Sanchez} {et~al.}(2016){Sanchez}, {Scoccimarro}, {Crocce}, {Grieb},
  {Salazar-Albornoz}, {DallaVecchia}, {Lippich}, {Beutler}, {Brownstein},
  {Chuang}, {Eisenstein}, {Kitaura}, {Olmstead}, {Percival}, {Prada},
  {Rodriguez-Torres}, {Ross}, {Samushia}, {Seo}, {Tinker}, {Tojeiro},
  {Vargas-Magana}, {Wang}, \& {Zhao}}]{Sanchez:2016aa}
{Sanchez}, A.~G., {Scoccimarro}, R., {Crocce}, M., {et~al.} 2016,
  \JournalTitle{ArXiv e-prints},
  \href{http://arxiv.org/abs/1607.03147}{{\sffamily arXiv:1607.03147}}

\bibitem[{{Sato} \& {Matsubara}(2011)}]{Sato:2011aa}
{Sato}, M., \& {Matsubara}, T. 2011,
  \href{http://dx.doi.org/10.1103/PhysRevD.84.043501}{\JournalTitle{\prd}, 84,
  043501}

\bibitem[{{Schlegel} {et~al.}(2011){Schlegel}, {Abdalla}, {Abraham}, {Ahn},
  {Allende Prieto}, {Annis}, {Aubourg}, {Azzaro}, {Baltay}, {Baugh}, {Bebek},
  {Becerril}, {Blanton}, {Bolton}, {Bromley}, {Cahn}, {Carton},
  {Cervantes-Cota}, {Chu}, {Cortes}, {Dawson}, {Dey}, {Dickinson}, {Diehl},
  {Doel}, {Ealet}, {Edelstein}, {Eppelle}, {Escoffier}, {Evrard}, {Faccioli},
  {Frenk}, {Geha}, {Gerdes}, {Gondolo}, {Gonzalez-Arroyo}, {Grossan},
  {Heckman}, {Heetderks}, {Ho}, {Honscheid}, {Huterer}, {Ilbert}, {Ivans},
  {Jelinsky}, {Jing}, {Joyce}, {Kennedy}, {Kent}, {Kieda}, {Kim}, {Kim},
  {Kneib}, {Kong}, {Kosowsky}, {Krishnan}, {Lahav}, {Lampton}, {LeBohec}, {Le
  Brun}, {Levi}, {Li}, {Liang}, {Lim}, {Lin}, {Linder}, {Lorenzon}, {de la
  Macorra}, {Magneville}, {Malina}, {Marinoni}, {Martinez}, {Majewski},
  {Matheson}, {McCloskey}, {McDonald}, {McKay}, {McMahon}, {Menard},
  {Miralda-Escude}, {Modjaz}, {Montero-Dorta}, {Morales}, {Mostek}, {Newman},
  {Nichol}, {Nugent}, {Olsen}, {Padmanabhan}, {Palanque-Delabrouille}, {Park},
  {Peacock}, {Percival}, {Perlmutter}, {Peroux}, {Petitjean}, {Prada},
  {Prieto}, {Prochaska}, {Reil}, {Rockosi}, {Roe}, {Rollinde}, {Roodman},
  {Ross}, {Rudnick}, {Ruhlmann-Kleider}, {Sanchez}, {Sawyer}, {Schimd},
  {Schubnell}, {Scoccimaro}, {Seljak}, {Seo}, {Sheldon}, {Sholl},
  {Shulte-Ladbeck}, {Slosar}, {Smith}, {Smoot}, {Springer}, {Stril}, {Szalay},
  {Tao}, {Tarle}, {Taylor}, {Tilquin}, {Tinker}, {Valdes}, {Wang}, {Wang},
  {Weaver}, {Weinberg}, {White}, {Wood-Vasey}, {Yang}, {Yeche}, {Zakamska},
  {Zentner}, {Zhai}, \& {Zhang}}]{Schlegel:2011aa}
{Schlegel}, D., {Abdalla}, F., {Abraham}, T., {et~al.} 2011,
  \JournalTitle{ArXiv e-prints},
  \href{http://arxiv.org/abs/1106.1706}{{\sffamily arXiv:1106.1706
  [astro-ph.IM]}}

\bibitem[{{Scoccimarro}(2015)}]{Scoccimarro:2015aa}
{Scoccimarro}, R. 2015,
  \href{http://dx.doi.org/10.1103/PhysRevD.92.083532}{\JournalTitle{\prd}, 92,
  083532}

\bibitem[{{Scoccimarro} \& {Sheth}(2002)}]{Scoccimarro:2002aa}
{Scoccimarro}, R., \& {Sheth}, R.~K. 2002,
  \href{http://dx.doi.org/10.1046/j.1365-8711.2002.04999.x}{\JournalTitle{\mnras},
  329, 629}

\bibitem[{{Sefusatti} {et~al.}(2016){Sefusatti}, {Crocce}, {Scoccimarro}, \&
  {Couchman}}]{Sefusatti:2016aa}
{Sefusatti}, E., {Crocce}, M., {Scoccimarro}, R., \& {Couchman}, H.~M.~P. 2016,
  \href{http://dx.doi.org/10.1093/mnras/stw1229}{\JournalTitle{\mnras}},
  \href{http://arxiv.org/abs/1512.07295}{{\sffamily arXiv:1512.07295}}

\bibitem[{{Springel}(2005)}]{Springel:2005aa}
{Springel}, V. 2005,
  \href{http://dx.doi.org/10.1111/j.1365-2966.2005.09655.x}{\JournalTitle{\mnras},
  364, 1105}

\bibitem[{{Sunayama} {et~al.}(2016){Sunayama}, {Padmanabhan}, {Heitmann},
  {Habib}, \& {Rangel}}]{Sunayama:2016aa}
{Sunayama}, T., {Padmanabhan}, N., {Heitmann}, K., {Habib}, S., \& {Rangel}, E.
  2016,
  \href{http://dx.doi.org/10.1088/1475-7516/2016/05/051}{\JournalTitle{\jcap},
  5, 051}

\bibitem[{{Takada} {et~al.}(2014){Takada}, {Ellis}, {Chiba}, {Greene},
  {Aihara}, {Arimoto}, {Bundy}, {Cohen}, {Dor{\'e}}, {Graves}, {Gunn},
  {Heckman}, {Hirata}, {Ho}, {Kneib}, {F{\`e}vre}, {Lin}, {More}, {Murayama},
  {Nagao}, {Ouchi}, {Seiffert}, {Silverman}, {Sodr{\'e}}, {Spergel}, {Strauss},
  {Sugai}, {Suto}, {Takami}, \& {Wyse}}]{Takada:2014aa}
{Takada}, M., {Ellis}, R.~S., {Chiba}, M., {et~al.} 2014,
  \href{http://dx.doi.org/10.1093/pasj/pst019}{\JournalTitle{\pasj}, 66, R1}

\bibitem[{{Taruya} {et~al.}(2012){Taruya}, {Bernardeau}, {Nishimichi}, \&
  {Codis}}]{Taruya:2012aa}
{Taruya}, A., {Bernardeau}, F., {Nishimichi}, T., \& {Codis}, S. 2012,
  \href{http://dx.doi.org/10.1103/PhysRevD.86.103528}{\JournalTitle{\prd}, 86,
  103528}

\bibitem[{{Taruya} {et~al.}(2014){Taruya}, {Koyama}, {Hiramatsu}, \&
  {Oka}}]{Taruya:2014aa}
{Taruya}, A., {Koyama}, K., {Hiramatsu}, T., \& {Oka}, A. 2014,
  \href{http://dx.doi.org/10.1103/PhysRevD.89.043509}{\JournalTitle{\prd}, 89,
  043509}

\bibitem[{{Taruya} {et~al.}(2013){Taruya}, {Nishimichi}, \&
  {Bernardeau}}]{Taruya:2013aa}
{Taruya}, A., {Nishimichi}, T., \& {Bernardeau}, F. 2013,
  \href{http://dx.doi.org/10.1103/PhysRevD.87.083509}{\JournalTitle{\prd}, 87,
  083509}

\bibitem[{{Taruya} {et~al.}(2010){Taruya}, {Nishimichi}, \&
  {Saito}}]{Taruya:2010aa}
{Taruya}, A., {Nishimichi}, T., \& {Saito}, S. 2010,
  \href{http://dx.doi.org/10.1103/PhysRevD.82.063522}{\JournalTitle{\prd}, 82,
  063522}

\bibitem[{{Tassev} {et~al.}(2015){Tassev}, {Eisenstein}, {Wandelt}, \&
  {Zaldarriaga}}]{Tassev:2015aa}
{Tassev}, S., {Eisenstein}, D.~J., {Wandelt}, B.~D., \& {Zaldarriaga}, M. 2015,
  \JournalTitle{ArXiv e-prints},
  \href{http://arxiv.org/abs/1502.07751}{{\sffamily arXiv:1502.07751}}

\bibitem[{{Tinker} {et~al.}(2012){Tinker}, {Sheldon}, {Wechsler}, {Becker},
  {Rozo}, {Zu}, {Weinberg}, {Zehavi}, {Blanton}, {Busha}, \&
  {Koester}}]{Tinker:2012aa}
{Tinker}, J.~L., {Sheldon}, E.~S., {Wechsler}, R.~H., {et~al.} 2012,
  \href{http://dx.doi.org/10.1088/0004-637X/745/1/16}{\JournalTitle{\apj}, 745,
  16}

\bibitem[{{White} {et~al.}(2014){White}, {Tinker}, \& {McBride}}]{White:2014aa}
{White}, M., {Tinker}, J.~L., \& {McBride}, C.~K. 2014,
  \href{http://dx.doi.org/10.1093/mnras/stt2071}{\JournalTitle{\mnras}, 437,
  2594}

\bibitem[{{Yan} {et~al.}(2004){Yan}, {White}, \& {Coil}}]{Yan:2004aa}
{Yan}, R., {White}, M., \& {Coil}, A.~L. 2004,
  \href{http://dx.doi.org/10.1086/383588}{\JournalTitle{\apj}, 607, 739}

\bibitem[{{Yoon} {et~al.}(2008){Yoon}, {Schawinski}, {Sheen}, {Ree}, \&
  {Yi}}]{Yoon:2008aa}
{Yoon}, J.~H., {Schawinski}, K., {Sheen}, Y.-K., {Ree}, C.~H., \& {Yi}, S.~K.
  2008, \href{http://dx.doi.org/10.1086/528958}{\JournalTitle{\apjs}, 176, 414}

\bibitem[{{Zehavi} {et~al.}(2002){Zehavi}, {Blanton}, {Frieman}, {Weinberg},
  {Mo}, {Strauss}, {Anderson}, {Annis}, {Bahcall}, {Bernardi}, {Briggs},
  {Brinkmann}, {Burles}, {Carey}, {Castander}, {Connolly}, {Csabai},
  {Dalcanton}, {Dodelson}, {Doi}, {Eisenstein}, {Evans}, {Finkbeiner},
  {Friedman}, {Fukugita}, {Gunn}, {Hennessy}, {Hindsley}, {Ivezi{\'c}}, {Kent},
  {Knapp}, {Kron}, {Kunszt}, {Lamb}, {Leger}, {Long}, {Loveday}, {Lupton},
  {McKay}, {Meiksin}, {Merrelli}, {Munn}, {Narayanan}, {Newcomb}, {Nichol},
  {Owen}, {Peoples}, {Pope}, {Rockosi}, {Schlegel}, {Schneider}, {Scoccimarro},
  {Sheth}, {Siegmund}, {Smee}, {Snir}, {Stebbins}, {Stoughton}, {SubbaRao},
  {Szalay}, {Szapudi}, {Tegmark}, {Tucker}, {Uomoto}, {Vanden Berk}, {Vogeley},
  {Waddell}, {Yanny}, \& {York}}]{Zehavi:2002aa}
{Zehavi}, I., {Blanton}, M.~R., {Frieman}, J.~A., {et~al.} 2002,
  \href{http://dx.doi.org/10.1086/339893}{\JournalTitle{\apj}, 571, 172}

\bibitem[{{Zehavi} {et~al.}(2005){Zehavi}, {Zheng}, {Weinberg}, {Frieman},
  {Berlind}, {Blanton}, {Scoccimarro}, {Sheth}, {Strauss}, {Kayo}, {Suto},
  {Fukugita}, {Nakamura}, {Bahcall}, {Brinkmann}, {Gunn}, {Hennessy},
  {Ivezi{\'c}}, {Knapp}, {Loveday}, {Meiksin}, {Schlegel}, {Schneider},
  {Szapudi}, {Tegmark}, {Vogeley}, {York}, \& {SDSS
  Collaboration}}]{Zehavi:2005aa}
{Zehavi}, I., {Zheng}, Z., {Weinberg}, D.~H., {et~al.} 2005,
  \href{http://dx.doi.org/10.1086/431891}{\JournalTitle{\apj}, 630, 1}

\bibitem[{{Zehavi} {et~al.}(2011){Zehavi}, {Zheng}, {Weinberg}, {Blanton},
  {Bahcall}, {Berlind}, {Brinkmann}, {Frieman}, {Gunn}, {Lupton}, {Nichol},
  {Percival}, {Schneider}, {Skibba}, {Strauss}, {Tegmark}, \&
  {York}}]{Zehavi:2011aa}
---. 2011,
  \href{http://dx.doi.org/10.1088/0004-637X/736/1/59}{\JournalTitle{\apj}, 736,
  59}

\bibitem[{{Zhao} {et~al.}(2013){Zhao}, {Saito}, {Percival}, {Ross},
  {Montesano}, {Viel}, {Schneider}, {Manera}, {Miralda-Escud{\'e}},
  {Palanque-Delabrouille}, {Ross}, {Samushia}, {S{\'a}nchez}, {Swanson},
  {Thomas}, {Tojeiro}, {Y{\`e}che}, \& {York}}]{Zhao:2013aa}
{Zhao}, G.-B., {Saito}, S., {Percival}, W.~J., {et~al.} 2013,
  \href{http://dx.doi.org/10.1093/mnras/stt1710}{\JournalTitle{\mnras}, 436,
  2038}

\end{thebibliography}

\appendix
\label{app:AppA}

For reference, here we list the first few polynomials $H_{l_> l_<}(x)$ from Eq.~(\ref{foffdiag})
\beqa
H_{20}(x)&=&x^2-1, \\ & & \nonumber \\
H_{40}(x)&=&{7\over 4}x^4-{5\over 2}x^2 +{3\over 4}, \\  & & \nonumber \\
H_{42}(x)&=&x^4-x^2, \\& & \nonumber \\
H_{60}(x)&=&	\frac{33}{8} x^6 - \frac{63}{8}x^4 + \frac{35}{8}x^2 - \frac{5}{8} 	, \\ & & \nonumber \\
H_{62}(x) &=&   \frac{11}{4}x^6 - \frac{9}{2}x^4 + \frac{7}{4}x^2, \\ & & \nonumber \\
H_{64}(x) &=&  x^6 -  x^4 
\label{Hpoly}
\eeqa

\end{document}